\documentclass{aastex631}
\usepackage{multirow}
\usepackage{wrapfig}

\usepackage{amsmath}
\usepackage{amssymb}
\usepackage{multirow}
\usepackage{graphicx} 
\usepackage{hhline}
\usepackage[T1]{fontenc}
\usepackage{courier}

\begin{document}

%%%%%%%%%%%%%%%%%%%%%%%%%%%%%%%%%
\title{Galaxy Phase-Space and Field-Level Cosmology: The Strength of Semi-Analytic Models}

%%%%%%%%%%%%%%%%%%%%%%%%%%%%%%%%

\correspondingauthor{Natalí S. M. de Santi}
\email{natalidesanti@gmail.com}

\author[0000-0002-4728-6881]{Natalí S. M. de Santi}
\affiliation{Berkeley Center for Cosmological Physics, University of California, Berkeley, 341 Campbell Hall, Berkeley, CA 94720, USA}
\affiliation{Physics Division, Lawrence Berkeley National Laboratory, 1 Cyclotron Road, Berkeley, CA 94720, USA}
\affiliation{Center for Computational Astrophysics, Flatiron Institute, 162 5th Avenue, New York, NY, 10010, USA}

\author[0000-0002-4816-0455]{Francisco Villaescusa-Navarro}
\affiliation{Center for Computational Astrophysics, Flatiron Institute, 162 5th Avenue, New York, NY, 10010, USA}
\affiliation{Department of Astrophysical Sciences, Princeton University, 4 Ivy Lane, Princeton, NJ 08544 USA}

\author[0000-0003-2860-5717]{Pablo Araya-Araya}
\affiliation{Cosmic Dawn Center (DAWN), Copenhagen, Denmark}
\affiliation{DTU Space, Technical University of Denmark, Elektrovej 327, DK2800 Kgs. Lyngby, Denmark}

\author{Gabriella De Lucia}
\affiliation{INAF - Astronomical Observatory of Trieste, via G.B. Tiepolo 11, I-34143 Trieste, Italy}
\affiliation{IFPU - Institute for Fundamental Physics of the Universe, via Beirut 2, 34151, Trieste, Italy}

\author[0000-0003-4744-0188]{Fabio Fontanot}
\affiliation{INAF - Astronomical Observatory of Trieste, via G.B. Tiepolo 11, I-34143 Trieste, Italy}
\affiliation{IFPU - Institute for Fundamental Physics of the Universe, via Beirut 2, 34151, Trieste, Italy}

\author[0000-0002-8449-1956]{Lucia A. Perez}
\affiliation{Center for Computational Astrophysics, Flatiron Institute, 162 5th Avenue, New York, NY, 10010, USA}
\affiliation{Department of Astrophysical Sciences, Princeton University, 4 Ivy Lane, Princeton, NJ 08544, USA}

\author[0009-0000-1302-3090]{Manuel Arnés-Curto}
\affiliation{Departamento de F\'isica Te\'orica,  Universidad Aut\'onoma de Madrid, 28049 Madrid, Spain}
\affiliation{Departamento de Física Fundamental and IUFFyM, Universidad de Salamanca, E-37008 Salamanca, Spain}

\author[0000-0001-9938-2755]{Violeta Gonzalez-Perez}
\affiliation{Departamento de F\'isica Te\'orica,  Universidad Aut\'onoma de Madrid, 28049 Madrid, Spain}
\affiliation{Centro de Investigación Avanzada en Física Fundamental (CIAFF), Facultad de Ciencias, Universidad Autónoma de Madrid, ES-28049 Madrid, Spain}

\author[0009-0004-1163-0160]{Ángel Chandro-Gómez}
\affiliation{Departamento de F\'isica Te\'orica,  Universidad Aut\'onoma de Madrid, 28049 Madrid, Spain}
\affiliation{International Centre for Radio Astronomy Research, The University of Western Australia, 35 Stirling Highway, Crawley, Western Australia 6009, Australia}
\affiliation{ARC Centre for All-Sky Astrophysics in 3 Dimensions (ASTRO 3D)}

\author[0000-0002-6748-6821]{Rachel S. Somerville}
\affiliation{Center for Computational Astrophysics, Flatiron Institute, 162 5th Avenue, New York, NY, 10010, USA}

\author[0000-0002-6292-3228]{Tiago Castro}
\affiliation{INAF-Osservatorio Astronomico di Trieste, Via G. B. Tiepolo 11, I-34143 Trieste, 
Italy}
\affiliation{INFN, Sezione di Trieste, Via Valerio 2, I-34127 Trieste TS, Italy}
\affiliation{IFPU, Institute for Fundamental Physics of the Universe, via Beirut 2, 34151 
Trieste, Italy}
\affiliation{CSC - Centro Nazionale di Ricerca in High Performance Computing, Big Data e Quantum Computing, Via Magnanelli 2, Bologna, Italy}

%%%%%%%%%%%%%%%%%%%%%%%%%%%%%%%%%%%
\begin{abstract}

Semi-analytic models are a widely used approach to simulate galaxy properties within a cosmological framework, relying on simplified yet physically motivated prescriptions.
They have also proven to be an efficient alternative for generating accurate galaxy catalogs, offering a faster and less computationally expensive option compared to full hydrodynamical simulations. 
In this paper, we demonstrate that using only galaxy $3$D positions and radial velocities, we can train a graph neural network coupled to a moment neural network to obtain a robust machine learning based model capable of estimating the matter density parameters, $\Omega_{\rm m}$, with a precision of approximately 10\%. 
The network is trained on ($25 h^{-1}$Mpc)$^3$ volumes of galaxy catalogs from L-Galaxies and can successfully extrapolate its predictions to other semi-analytic models (GAEA, SC-SAM, and {\sc Shark}) and, more remarkably, to hydrodynamical simulations (\texttt{Astrid}, {\sc SIMBA}, IllustrisTNG, and SWIFT-EAGLE). 
Our results show that the network is robust to variations in astrophysical and subgrid physics, cosmological and astrophysical parameters, and the different halo-profile treatments used across simulations.
This suggests that the physical relationships encoded in the phase-space of semi-analytic models are largely independent of their specific physical prescriptions, reinforcing their potential as tools for the generation of realistic mock catalogs for cosmological parameter inference.

\end{abstract}

%%%%%%%%%%%%%%%%%%%%%%%%%%%%%%%%%%%%%%

%%%%%%%%%%%%%%%%%%%%%%%%%%%%%%%%%%%%%%
\section{Introduction} 
\label{sec:intro}
%%%%%%%%%%%%%%%%%%%%%%%%%%%%%%%%%%%%%%

It is well known that galaxies are not randomly distributed in space; rather, they are observables that trace the cosmic web, the large scale structure of the Universe, where dark matter (DM) plays a central role in structure formation. 
Galaxies not only trace the underlying DM distribution but also maintain an intricate relationship with the DM halos in which they form and evolve, what is commonly referred to as the halo-galaxy connection.
Modeling galaxy evolution within a cosmological context therefore represents one of the greatest challenges in astrophysics, owing to the vast range of scales and the multitude of physical processes involved \citep{Somerville2015}.

Over the past decades, various methods have been developed to model the physics underlying the halo-galaxy connection.
These approaches generally fall into two main categories: physical and empirical models \citep{Wechsler2018, DeLucia2019}.
Physical models, including hydrodynamical simulations and semi-analytic models (SAMs), aim to capture the fundamental processes that govern galaxy formation within DM halos.
In contrast, empirical models such as subhalo abundance matching \citep[SHAM, e.g.][]{kravtsov2004,yu2024} and halo occupation distribution \citep[HOD, e.g.][]{zheng2005,avila2020} frameworks are more data-driven, relying on statistical correlations between galaxy and halo properties inferred from simulations or observations.

In this work, we focus on the two principal physics based techniques that serve as our datasets: hydrodynamical simulations and SAMs.
This choice is motivated by the fact both methods produce remarkably consistent predictions, that are in good agreement with observations \citep{Somerville2015, Wechsler2018}.
Furthermore, using a range of models allow us to account for the complex physics of the halo-galaxy connection and to mitigate the impact of differences between modeling approaches.

Hydrodynamical simulations solve the equations of gravity, hydrodynamics, and thermodynamics using numerical techniques applied to particles and/or grid cells representing DM and gas \citep{Somerville2015, Crain2023}.
Their main limitations, however, stem from their high computational cost and from the still incomplete understanding of the small-scale physics governing baryonic processes, which must be incorporated through various subgrid prescriptions.

In this context, SAMs represent one of the most effective current alternatives for predicting observables in the Universe \citep{Baugh2006, Benson2010, Somerville2015, Gonzalez-Perez2020, Gomez2025}.
They are built on top of DM-only simulations, in which various physical processes are approximated through analytic prescriptions that can be tracked along the merger histories of DM halos \citep{Somerville1999, Hirschmann2016, Henriques2020}.
As a result, SAMs drastically reduce computational demands compared to fully numerical hydrodynamical simulations, enabling both the efficient exploration of a wide parameter space and the generation of catalogs over much larger volumes.
Their main limitation, however, lies in the approximations themselves: they are less capable of capturing small-scale physical effects and can differ in their assumptions for the underlying halo profile and gas dynamic prescriptions.

Recent studies \citep{Anagnostidis_2022, pablo-galaxies-2022, helen-halos-2022, lucas2022, deSanti2023-1, deSanti2023-2, Shao2023-2, Roncoli2023, Chatterjee2024, Cuesta-Lazaro2024} have demonstrated that a powerful approach to constraining cosmological parameters from galaxy and/or halo catalogs involves representing them as point clouds or graphs, and performing field-level simulation-based inference (SBI), also referred to as likelihood-free inference (LFI) or implicit likelihood inference (ILI).
These works have primarily relied on DM-only or hydrodynamical simulations, leveraging their exact numerical solutions, which provide precise positions and velocities of the simulated structures.

Building upon the use of these methods for cosmological parameter inference, a key challenge, especially when aiming for applications to real data, is the development of robust machine learning (ML) suites.
These models are inherently data-driven, meaning they learn patterns based on the intrinsic properties of the data they are trained on and, consequently, often fail to extrapolate accurately to datasets that differ from their training domain \citep{robustness-Hassani2022}.
Since we do not yet know which physical model best describes our Universe, it is essential to design methods that exhibit consistent predictive power across different simulation frameworks, and then more importantly, are able to extract cosmological and astrophysical information once applied to real observations.

The lack of robustness in current ML models remains an open issue and can be attributed to several factors:
(1) limited overlap between datasets, arising from differences in the chosen astrophysical parameter variations and in the resulting galaxy-property distributions;
(2) models learning spurious or non-physical features (e.g., numerical artifacts); and
(3) differences in data representation, such as comparing raw phase-space information with compressed latent or manifold-based representations.

Robustness has previously been explored across the CAMELS hydrodynamical simulations using a variety of data formats, including 2D projected maps with convolutional neural networks \citep{Paco2021}, galaxy tabular data \citep{one_gal-2022}, neutral hydrogen maps \citep{Jo2025}, and galaxy and/or halo catalogs represented as graphs or point clouds \citep{pablo-galaxies-2022, helen-halos-2022, Shao2023-2, deSanti2023-1}.
Additionally, analyses using different N-body simulations were conducted by \cite{Bayer2025}, who showed that statistical out-of-distribution behavior can arise from both resolution effects and the sensitivity of ML methods to small-scale fluctuations.

Regarding galaxy catalogs, several approaches have already been explored to mitigate this lack of robustness.
One of them relies on domain adaptation (DA) techniques \citep{Csurka2017, Wang2018, Farahani2020, Kumari2023}, which aim to create models that generalize across different data domains. 
Most of these methods focus on modifying the loss function to include a measure of the discrepancy between the latent-space probability distributions of the different datasets.

DA techniques, when combined with graph neural networks (GNNs), have already been applied to the estimation of $\Omega_{\rm m}$ across different hydrodynamical simulations, specifically IllustrisTNG and {\sc SIMBA}, using the galaxies' phase-space information \citep{Roncoli2023}. 
The authors found that performance improved when training on {\sc SIMBA} and testing on IllustrisTNG, although a bias persisted.

An alternative strategy is to train ML models on datasets that are sufficiently diverse to encompass the variability present across different simulations.
This approach was adopted in \cite{deSanti2023-1}, where a GNN coupled with a moment neural network (MNN), trained on \texttt{Astrid} using galaxy phase-space information, was shown to successfully extrapolate its predictions of $\Omega_{\rm m}$ across five different subgrid physical models.
In that case, the model’s success was attributed to the greater diversity of the training dataset, both in the total number of galaxies per catalog and in the variation of their properties \citep{Ni2023}, combined with the chosen graph construction and GNN architecture.

Beyond assessing the robustness of different ML approaches, it is essential to build datasets using methods that are computationally fast and scalable.
In \cite{deSanti2023-1}, the first robust field-level model across multiple hydrodynamical simulations was demonstrated.
However, extending this idea by running thousands of large-volume hydrodynamical simulations, such as \texttt{Astrid}, is simply not feasible with current computational resources.
This is precisely why SAMs are so promising: they offer a rapid yet physically motivated way to generate diverse galaxy catalogs, making them strong candidates to match the performance of hydrodynamical simulations for SBI and, ultimately, for the creation of realistic mocks needed for real-data applications.

In this paper we make use of galaxy $3$D positions and radial velocities from L-Galaxies to train a GNN associated with a MNN to predict $\Omega_m$, looking for a robust model that could overcome the main differences over different SAMs and different hydrodynamic simulations. 
The manuscript is structured as follows: 
in Section \ref{sec:data} we describe the dataset used to train and test the model (specifications related to the SAMs are provided in Appendix \ref{sec:SAM+}); 
in Section \ref{sec:methodology} we describe the methodology, presenting an overview of the ML model used as well as the training procedure (more details can be found in the Appendix \ref{sec:MLdetails}); 
in Section \ref{sec:results} the results are shown and explained; 
Section \ref{sec:conclusions} presents a discussion and conclusions. 
Additionally, a Normalizing Flows (NFs) were coupled with the GNN instead of the MNN, to perform the full posterior inferences. 
The results of this experiment are shown and discussed in  Appendix \ref{sec:NF}.

%%%%%%%%%%%%%%%%%%%%%%%%%%%%%%%%%%%%%%%%%%%%
\section{Data} 
\label{sec:data}
%%%%%%%%%%%%%%%%%%%%%%%%%%%%%%%%%%%%%%%%%%%%

%%%%%%%%%%%%%%%%%%%%%%%%%%%%%%%%%%%%%%%%%%%
\subsection{Simulations} 
\label{sec:simulations}

\begin{table*}[h!]
 \caption{\label{tab:summary} Characteristics of the SAMs and hydrodynamic simulations used in this work.}
 \begin{center}
% \resizebox{\textwidth}{!}{%
  \begin{tabular}{cccccc}
   \hline\hline
   \multirow{3}{*}{\textbf{Model}} & \multirow{3}{*}{\textbf{Type}} & \multirow{3}{*}{\textbf{Usage}} & 
   {\bf Number of} & \textbf{Mean number} & \multirow{3}{*}{{\bf Reference}} \\
   & & & {\bf simulations} & \textbf{of galaxies} & \\
   & & & {\bf used} & \textbf{per catalog} & \\
   \hline\hline
    \multirow{2}{*}{L-Galaxies} & \multirow{2}{*}{SAM} & Train, validate & \multirow{2}{*}{$1,000$(LH) + $27$(CV)} & \multirow{2}{*}{$5,233$} & \multirow{2}{*}{\cite{Henriques2020}}\\
    & & \& Test & & & \\
    GAEA & SAM & Test & $1,000$(LH) + $27$(CV) & $1,868$ & 
    \cite{DeLucia2024} \\
    SC-SAM & SAM & Test & $50$(LH) + $23$(CV) & $1,337$ & 
    \cite{Perez2023} \\
    {\sc Shark} & SAM & Test & $1,000$(LH) + $27$(CV) & $940$ & 
    \cite{Lagos2024} \\
    \texttt{Astrid} & Hydro & Test & $1,000$(LH) + $27$(CV) & $1,114$ & 
    \cite{Astrid2022} \\
    {\sc SIMBA} & Hydro & Test & $1,000$(LH) + $27$(CV) & $1,093$ & 
    \cite{SIMBA2019}                 \\
    IllustrisTNG & Hydro & Test & $1,000$(LH) + $27$(CV) + $1,024$(SB) & $737$  &
    \cite{Pillepich2018}             \\
    Magneticum & Hydro & Test & $50$(LH) + $27$(CV)   & $3,655$ & \cite{MAGNETICUM2014}            \\
    SWIFT-EAGLE & Hydro & Test          & $1,000$(LH) + $27$(CV) & $1,220$ & \cite{EAGLE2015}            \\
   \cline{1-6}
   \end{tabular}%}
  \end{center}
\end{table*}

The galaxy catalogs we use to train, validate, and test our models come from thousands of hydrodynamic and semi-analytic simulations of the Cosmology and Astrophysics with MachinE Learning Simulations -- CAMELS project \citep{Paco2021-projCAMELS, Villaescusa-Navarro2022-relCAMELS}, that include periodic boxes of $25~h^{-1}{\rm Mpc}$ on a side. 
All the simulations follow the evolution of $256^3$ DM particles and are initialized with $256^3$ fluid elements from $z = 127$ down to $z = 0$.
The catalogs used in this work correspond to $z = 0$.

The CAMELS simulations can be classified into different sets and suites depending on how 
their parameters are arranged and which code was used to run them. We start by classifying 
the catalogs into different sets:
\begin{itemize}
 \item \textbf{Latin Hypercube (LH).} The simulations in this category have their cosmological and astrophysical parameters sampled according to a LH design.
For the cosmological parameters, their values span the following ranges: $\Omega_{\rm m} \in [0.1, 0.5]$ and $\sigma_8 \in [0.6, 1.0]$.
In the hydrodynamical simulations, only four astrophysical parameters are varied: $A_{\rm SN1}$, $A_{\rm SN2}$, $A_{\rm AGN1}$, and $A_{\rm AGN2}$.
These parameters control the efficiency of supernova (SN) and active galactic nuclei (AGN) feedback, although their specific definitions vary across the different hydrodynamic models.
For explicit details, see \citet{Paco2021-projCAMELS, Ni2023}.
The parameter ranges are: $A_{\rm SN1} \in [0.25, 4.0]$, $A_{\rm SN2} \in [0.5, 2.0]$, $A_{\rm AGN1} \in [0.25, 4.0]$, and $A_{\rm AGN2} \in [0.5, 2.0]$.
For the SAMs, a much larger number of parameters are varied, and they are different depending on the SAM in question (due to their different physical prescriptions); these parameters are detailed in Appendix \ref{sec:SAM+}.
Other cosmological parameters such as $\Omega_{\rm b}$, $h$, $n_s$, and $w$ follow the same values of the Cosmic Variance set.
Each of the simulations in the LH has been run with a different initial random seed for the generation of the initial conditions. We used these simulations for training, validating, and testing the ML model.
 
 \item {\bf Cosmic Variance (CV).} These simulations have been run with the fiducial value of the cosmological and astrophysical parameters, that follows for: $\Omega_m = 0.3$, $\sigma_8 = 0.8$, $\Omega_{\rm b} = 0.049$, $h = 0.6711$, $n_s = 0.9624$, $w = - 1$, 
$M_\nu = 0$ eV and $A_{SN1} = A_{SN2} = A_{AGN1} = A_{AGN2} = 1$.
 The fiducial values employed for each SAM can also be found in Appendix \ref{sec:SAM+}.
 The initial conditions for each simulation in this set  have been generated with a different initial random seed. These simulations are only used for testing the ML models.

 \item \textbf{Sobol Sequence (SB).} The simulations in this set have their cosmological and astrophysical parameters arranged following a Sobol sequence \citep{SOBOL1967}.
A total of $28$ parameters are varied: five cosmological ($\Omega_{\rm m}$, $\Omega_{\rm b}$, $h$, $n_s$, $\sigma_8$) and $23$ astrophysical.
The astrophysical parameters include the commonly used ones ($A_{\rm SN1}$, $A_{\rm SN2}$, $A_{\rm AGN1}$, $A_{\rm AGN2}$) as well as others related to star formation, galactic winds, black hole (BH) growth, and quasar activity.
All parameters vary within ranges centered around their fiducial values.
We use these simulations only for testing \citep[and only under the IllustrisTNG prescription accordingly to][]{Ni2023} to assess how well our ML models generalize to the new effects introduced by these parameter variations, which differ from those explored during training.
\end{itemize}

The SAMs have been run over DM halo merger trees, generated by the DM-only counterpart from the \texttt{Astrid} hydrodynamic simulation, ran using MP-Gadget \citep{MPGadget}.
The merger trees have been obtained with {\sc SubLink} \citep{Nelson2015, Rodriguez-Gomez2015}, using their extended format and halos found with \textsc{SubFind} \citep{Springel2001, Dolag2009}, for L-Galaxies and GAEA. 
SC-SAM and Shark were run with {\sc Rockstar} subhalo catalogs \citep{Behroozi2013a} in combination with {\sc ConsistentTrees} \citep{Behroozi2013b}.
All these merger trees follow the same particle and redshift evolution, for the same size boxes and cosmological parameters of the former simulations described.

The CAMELS-SAM simulations can also be classified into different model suites according to the code used to run them. See Appendix \ref{sec:SAM+} for more details about the parameters that are varied:

\begin{itemize}

\item {\bf L-Galaxies.} This SAM was run on the DM-only counterpart from the \texttt{Astrid} simulations, using merger trees obtained using {\sc SubLink} and subhalos from \textsc{SubFind} and adopts the galaxy formation prescriptions from the \citet{Henriques2020} version of L-Galaxies\footnote{\url{https://lgalaxiespublicrelease.github.io/}}. 
The positions and velocities of satellite galaxies are normally tracked using the most bound DM particle, obtained from an input file optimized for the Millennium simulation \citep{Springel2005}. 
However, because generating this input is not straightforward for the \texttt{Astrid} simulation, these quantities are instead extrapolated from the properties of the last subhalo before it becomes disrupted. 
The \citet{Henriques2020} model was calibrated using a Markov Chain Monte Carlo (MCMC) approach, constrained by the stellar mass function and the fraction of passive (quiescent) galaxies at $z = 0$ and $z = 2$, as well as by the neutral hydrogen mass function at $z = 0$. L-Galaxies includes $19$ free parameters in total; following the calibration framework of \citet{Henriques2020}, we varied $14$ of them\footnote{\citet{Henriques2020} also calibrated an additional parameter related to ram pressure stripping, which we fixed at their best-fit value.}. 
These parameters govern the equations describing key astrophysical processes, including secular star formation, merger-induced starbursts, supernova feedback, the growth of supermassive BHs and their associated AGN feedback, the reincorporation timescale of ejected gas, and tidal disruption. 
The $14$ free parameters, their corresponding variation ranges, and fiducial values are listed in Table \ref{tab:LGalaxies_parameters}. 
They were sampled using the LH algorithm, generating $1,000$ LH catalogs, while the default \citet{Henriques2020} values were used to produce $27$ control (CV) catalogs.

 \item {\bf GAEA.} The GAEA\footnote{An introduction to GAEA, a list of recent work, as well as a data file containing published model predictions, can be found at \href{https://sites.google.com/inaf.it/gaea/home}{https://sites.google.com/inaf.it/gaea/home}.} (GAlaxy Evolution and Assembly) SAM is based on \cite{DeLucia2014, Hirschmann2016, Xie2017, Fontanot2017, Fontanot2020, Xie2020, DeLucia2024, Xie2024} and their boxes were run over merger trees obtained with {\sc SubLink} and subhalos with \textsc{SubFind} from the DM counterpart of \texttt{Astrid}.
 In GAEA, galaxy positions and velocities are treated as follows:
 central galaxies are placed at the center of the most bound subhalo and inherit its velocity.
 Satellite galaxies are associated with distinct DM subhalos identified with {\sc SubFind} \citep{Springel2001, Dolag2009}, and their positions and velocities correspond to those of their host subhalo.
 When a satellite's subhalo is stripped below the resolution limit of the simulation, the galaxy becomes an orphan and its position and velocity are assigned to those of the most bound particle at the last snapshot in which the subhalo could still be identified \citep{Xie2020}.
 From the free parameters available, four of them were varied in a LH and used to run $1,000$ LH catalogs: two related to the SN efficiency and two related to AGN feedback. 
 A more detailed explanation about these parameters can be found in Table \ref{tab:gaea_parameters}, as well as their varied range and fiducial values.
 The model has been formally calibrated on the Millennium simulation \citep{Springel2005} and their parameters have been used to produce $27$ CV boxes \citep{DeLucia2024}. Additionally, the same set of parameters has been proved to be robust at different Millennium resolutions \citep{Fontanot2025}.

 \item {\bf SC-SAM.} These SAMs have been run over DM-only merger trees from \texttt{Astrid}, obtained from {\sc Rockstar} catalogs in combination with {\sc ConsistentTrees}, and based on the model presented in \cite{Somerville2008, Somerville_sam2015}. Three free parameters ($A_{\rm SN1}$, multiplicative pre-factor to $\epsilon_{\rm SN}$; $A_{\rm SN2}$, additive pre-factor to $\alpha_{rh}$; and $A_{\rm AGN}$, multiplicative pre-factor to $\kappa_{\rm radio}$;  \citealt{Perez2023}) have been varied to produce $50$ LH catalogs (see Table \ref{tab:S-SAM_parameters} for more details, ranges and fiducial values). 
 The orbits of satellite galaxies are tracked using an analytic dynamical friction model \citep{Somerville2008}, and satellite positions are re-assigned in post-processing to follow an NFW profile \citep{NFW1997}.
 The model has been calibrated in order to follow observable relations such as stellar mass function, the stellar mass-halo mass relation, the cold gas fraction vs. stellar mass, stellar metallicity-stellar mass, and BH mass-bulge mass relationships accordingly to observations and IllustrisTNG300 \citep{Gabrielpillai2022, Perez2023}, producing $23$ CV catalogs.
 
 \item \textbf{\textsc{Shark}.} Catalogs of model galaxies from {\sc Shark} are based on the fiducial calibration presented in \cite{Lagos2018, Lagos2024}, with the exception of $v_{\rm hot}$, $\nu_{SF}$, e$_{\rm sb}$ and $\epsilon_{\rm disc}$. 
 These four parameters have been set to match different stellar mass function datasets (\citealt{Baldry_2012}, \citealt{Moustakas_2013}, \citealt{Ilbert_2013}, \citealt{LiWhite_2009}) when using halos from the CV boxes. 
 Hence, these parameters are slightly different from the fiducial ones. 
 The fiducial calibration was performed using merger trees from SURFS DM-only simulations \citep{Elahi2018}, with halos identified with the {\sc VELOCIraptor} finder \citep{velociraptop1, velociraptor2} and links made with {\sc TreeFrog} \citep{Elahi2019}. 
 {\sc Shark} tracks satellite galaxies using subhalo information whenever the subhalo is resolved. 
 Once the subhalo disappears, the resulting orphan galaxies are followed analytically using a NFW-based prescription for their positions and velocities \citep{NFW1997}. 
 These model has been run over \texttt{Astrid}'s DM-only simulations also using {\sc Rockstar} catalogs in combination with {\sc ConsistentTrees}, although post-processed in this case by the code {\sc DHalo} \citep{Jiang2014}. 
 This model was used to generate $27$ CV boxes, sampling $16$ free parameters over $1,000$ LH simulations. 
 These model free parameters govern astrophysical processes such as stellar and AGN feedbacks, star formation, BH growth, gas reincorporation, tidal stripping and disc instabilities. 
 A summary of the parameters and their fiducial values and corresponding ranges is provided in Table \ref{tab:shark_parameters}.
\end{itemize}

We stress that the free parameters varied across the different SAMs do not follow a common pattern.
For instance, L-Galaxies and {\sc Shark} vary $14$ and $16$ parameters, respectively, whereas GAEA and SAM-SC vary only four and three.
Such differences naturally lead to variations in their resulting galaxy populations.
We stress that the free parameters varied over the different SAMs do not follow a pattern.

Although the SAMs presented in this paper are all built on the same \texttt{Astrid} DM-only counterpart, they have been calibrated and run using different halo finders and merger tree builders (e.g., {\sc SubLink}+{\sc SubFind}, {\sc RockStar}+{\sc ConsistentTrees}, and {\sc VELOCIraptor}+{\sc TreeFrog}).
Such differences can introduce significant scatter in galaxy properties \citep{Knebe2015}.
For example, \cite{Avila2014} showed that variations in halo-finder algorithms can substantially alter the mass-assembly histories from which merger trees are constructed.
Similarly, even when the same SAM is applied to different merger trees, large discrepancies in galaxy predictions may arise \citep{Lee2014}, although other studies have found good consistency in some cases \citep{Gomez2022}.
In addition, numerical artifacts inherent to each merger tree algorithm can further affect SAM predictions \citep{Angel2025}, reflecting both differences in merger tree construction \citep{Srisawat2013} and the halo finders on which they rely \citep{Avila2014}.
Finally, although beyond the scope of this work, different N-body codes can yield noticeably different DM fields that also require marginalization \citep{Bayer2025}, potentially compounding these effects.
% Although the SAMs presented in this paper have the same DM-only counterpart from Astrid, the different models have been calibrated and run using different halo finders/tree builders ({\sc SubLink}+{\sc SubFind}, {\sc RockStar}+{\sc ConsistentTrees}, and {\sc VELOCIraptor}+{\sc TreeFrog}).
% This can, firstly, introduce scatter in galaxy properties \citep{Knebe2015} -- as shown in \cite{Avila2014}, differences in halo finder codes can strongly change the mass histories that merger trees codes construct from the same input halo catalogs. 
% Secondly, even when the same SAM is applied to different merger trees that have been independently calibrated, substantial discrepancies can arise \citep{Lee2014}, although other studies find that galaxy properties remain consistent \citep{Gomez2022}. 
% Moreover, numerical artifacts inherent to each merger tree algorithm may also affect the SAM predictions \citep{Angel2025}, given both differences in merger tree algorithms \citep{Srisawat2013} as well as major differences due to different halo finders \citep{Avila2014}. 
% Finally, though beyond the scope of this work, N-body codes themselves can yield notably different DM fields that must be marginalized over \citep{Bayer2025} and might confound with these other phenomena as well.

The hydrodynamic simulations have been run with different codes that solve the hydrodynamic equations differently and implement different subgrid models: IllustrisTNG \citep{Weinberger2016, Pillepich2018}, {\sc SIMBA} \citep{SIMBA2019}, \texttt{Astrid} \citep{Astrid2022}, Magneticum \citep{MAGNETICUM2014}, and SWIFT-EAGLE \citep{EAGLE2015, Crain2015, Schaller2016, Schaller2024}. 

The CAMELS hydrodynamical simulations can also be classified into different model suites according to the 
code used to run them:

\begin{itemize}
 
 \item \textbf{\texttt{Astrid}.} These simulations were run using MP-Gadget \citep{MPGadget} applying some modifications to the subgrid model employed in the \texttt{Astrid} simulation 
 \citep{Astrid-Y-2022, Astrid2022, Ni2023}.  
 This suite contains $1,000$ LH and $27$ CV simulations.

\item \textbf{\textsc{SIMBA}.} These simulations were run with the \textsc{Gizmo} code \citep{Hopkins2015} and employ the same subgrid physics as the {\sc SIMBA} simulation \citep{SIMBA2019}. 
This suite contains $1,000$ LH and $27$ CV simulations.
 
 \item {\bf IllustrisTNG.} These simulations were run using \textsc{Arepo} 
 \citep{springel2010, Weinberger2020} applying the same subgrid physics as the IllustrisTNG simulations \citep{Weinberger2016, Pillepich2018}. 
 This suite 
 contains $1,000$ LH, $27$ CV, and $2048$ SB simulations. 
   
 \item {\bf Magneticum.} These simulations were run with the parallel cosmological Tree-PM code P-Gadget3 \citep{GADGET2}. 
 The code uses an entropy-conserving formulation of Smoothed Particle Hydrodynamics (SPH) \citep{Springel2002}, with SPH modifications according to  \cite{Dolag2004ApJ, Dolag2005MNRAS, Dolag2006MNRAS}. 
 It also includes prescriptions for multiphase interstellar medium based on the model by \cite{Springel2003MNRAS} as well as \cite{Tornatore2007} for the metal enrichment prescription. 
 The model follows the growth and evolution of BHs and their associated AGN feedback based on the model presented by \citet{Springel2005MNRAS} and \citet{DiMatteo2005}, but includes modifications based on  \citet{Fabjan2011MNRAS}, \citet{Hirschmann2014MNRAS} and \citet{Steinborn2016MNRAS}. 
 The set contains $50$ LH and $27$ CV simulations. 
 
\item {\bf SWIFT-EAGLE.} These simulations have been run with the \textsc{Swift} code \citep{Schaller2016, Schaller2018, Schaller2024} using a new subgrid physics model based on the original Gadget-EAGLE simulations \citep{EAGLE2015, Crain2015}, with some parameter changes \citep{EAGLE-Borrow2022}. 
The full model will be described in \cite{Lovell-prep}. 
This suite contains $1,000$ LH and $27$ CV simulations.

\end{itemize}

Finally, the halos and subhalos are identified in the hydrodynamical simulations for every snapshot using the halo and subhalo finder: \textsc{SubFind} \citep{Springel2001, Dolag2009}.

%%%%%%%%%%%%%%%%%%%%%%
\subsection{Galaxy catalogs}

In this work, we consider only galaxies with stellar masses above $1.3 \times 10^8\ M_\odot/h$, a threshold that lies well above the resolution and convergence limits of all simulations employed.
It is important to note that stellar masses are defined differently in SAMs and hydrodynamical simulations.
In SAMs, stellar masses arise from analytic prescriptions (e.g., {\sc Shark} separates bulge and disk components, while SAM-SC and GAEA track the total stellar content analytically).
In contrast, hydrodynamical simulations compute stellar masses directly from the numerical evolution, by summing the masses of all star particles bound to each galaxy's subhalo.

Each galaxy catalog is then constructed by selecting all galaxies whose stellar masses exceed the chosen threshold, and for every simulation we generate multiple catalogs by varying this mass limit.

A summary of the simulation characteristics is provided in Table \ref{tab:summary}, which lists if they are SAMs or hydrodynamic simulations, their respective usages (train, validation and test) for the ML model, the number of catalogs for the different sets (LH, SB, and CV), the mean number of galaxies per catalog\footnote{See Section \ref{sec:results} for a discussion regarding the different number of galaxies in the different simulations.}, and the reference for each of the original galaxy formation models.

%%%%%%%%%%%%%%%%%%%%%%%%%%%%%%%%%%%%
\section{Methodology}
\label{sec:methodology}
%%%%%%%%%%%%%%%%%%%%%%%%%%%%%%%%%%%%

In this section we describe: the method we use to construct graphs from galaxy catalogs (Section \ref{sec:the_graph}); the architecture of our GNN associated to the MNNs (Section \ref{sec:architecture}) and the training procedure and optimization choices (Section \ref{sec:train_optm}).
Any further detail related to this section can be found in Appendix \ref{sec:MLdetails}.

%%%%%%%%%%%%%%%%%%%%%%%%%%%%%%%
\subsection{Galaxy graphs}
\label{sec:the_graph}

The input for our GNNs are graphs: mathematical structures characterized by nodes, edges, and global properties. 
Every element of the graph can be described by a set of properties: $\mathbf{n}_i$ represents the properties of node $i$, $\mathbf{e}_{i j}$ represents the features of the edge between node $i$ and $j$, and $\mathbf{g}$ contains global properties of the graph \citep{Gilmer2017, Zhou2018, Battaglia2018}. 

We construct graphs from galaxy catalogs that contain the galaxy positions and their peculiar velocities (only the $z$ component), following the same prescription as used in \cite{deSanti2023-1}. 
Galaxies represent the graph nodes and two galaxies are connected by an edge if their distance is smaller than a given linking radius $r_{link}$, a value which we consider as a hyperparameter for the network.
We use as a global property of the graph the logarithm of the number of galaxies 
in the graph: $\log_{10} ( N_g )$.
The edge features contain information about the spatial distribution of galaxies (their positions), and those properties are designed to make the graph invariant under rotations and translations.
The exact node and edge attributes are described in \ref{sec:graph_details}.

%%%%%%%%%%%%%%%%%%%%%%%%%%%%%%%
\subsection{Architecture}
\label{sec:architecture}

The architecture we employ in this work also follows the one presented in \cite{deSanti2023-1} and \cite{pablo-galaxies-2022}\footnote{See \textsc{CosmoGraphNet} github repository: \href{https://github.com/PabloVD/CosmoGraphNet}{https://github.com/PabloVD/CosmoGraphNet}.}.
Basically, the GNN is trained to infer another global property of each input graph, the value of $\Omega_{\rm m}$ from each galaxy catalog. 
The main idea is to perform a transformation of the graph components' information (i.e.\ nodes $\mathbf{n}_i$, edges $\mathbf{e}_{i j}$, and global $\mathbf{g}$ attributes are updated), while the graph structure is preserved. 
This is known as message passing scheme (see the exact implementation in Section \ref{sec:architecture_details}).
In the end, the compressed graph information is converted by a multi-layer 
perceptron (MLP) into the final predicted property of the graph ($\Omega_\textrm{m}$). 
The loss function is designed so that the MNNs infer the first ($\mu$, see Equation \ref{eq:mu}) and second moments ($\sigma$, see Equation \ref{eq:sigma}) of the posterior distribution.
The number of layers, the number of neurons per layer, the weight decay, and the learning rate were considered as hyperparameters. The implementation of all the architecture presented in this work was done using \textsc{PyTorch Geometric} \citep{pytorch-geometric}.

\begin{table*}[h!]
 \caption{\label{tab:hyperparameters} Hyperparameters values selected for the best model.}
 \begin{center}
% \resizebox{\textwidth}{!}{%
  \begin{tabular}{cc}
   \hline\hline
   \textbf{Hyperparameter} & \textbf{Best value} \\
   \hline\hline
   Hidden Features & 27\\
   Number of GNN layers & 2\\
    $r_{link}$ & $1.11$Mpc/h\\
    Learning Rate & $2.13 \cdot 10^{-7}$\\
    Weight Decay & $9.5 \cdot 10^{-3}$\\
   \cline{1-2}
   \end{tabular}%}
  \end{center}
\end{table*}

\begin{figure}[h!]
 \centering
 \includegraphics[scale=0.35]{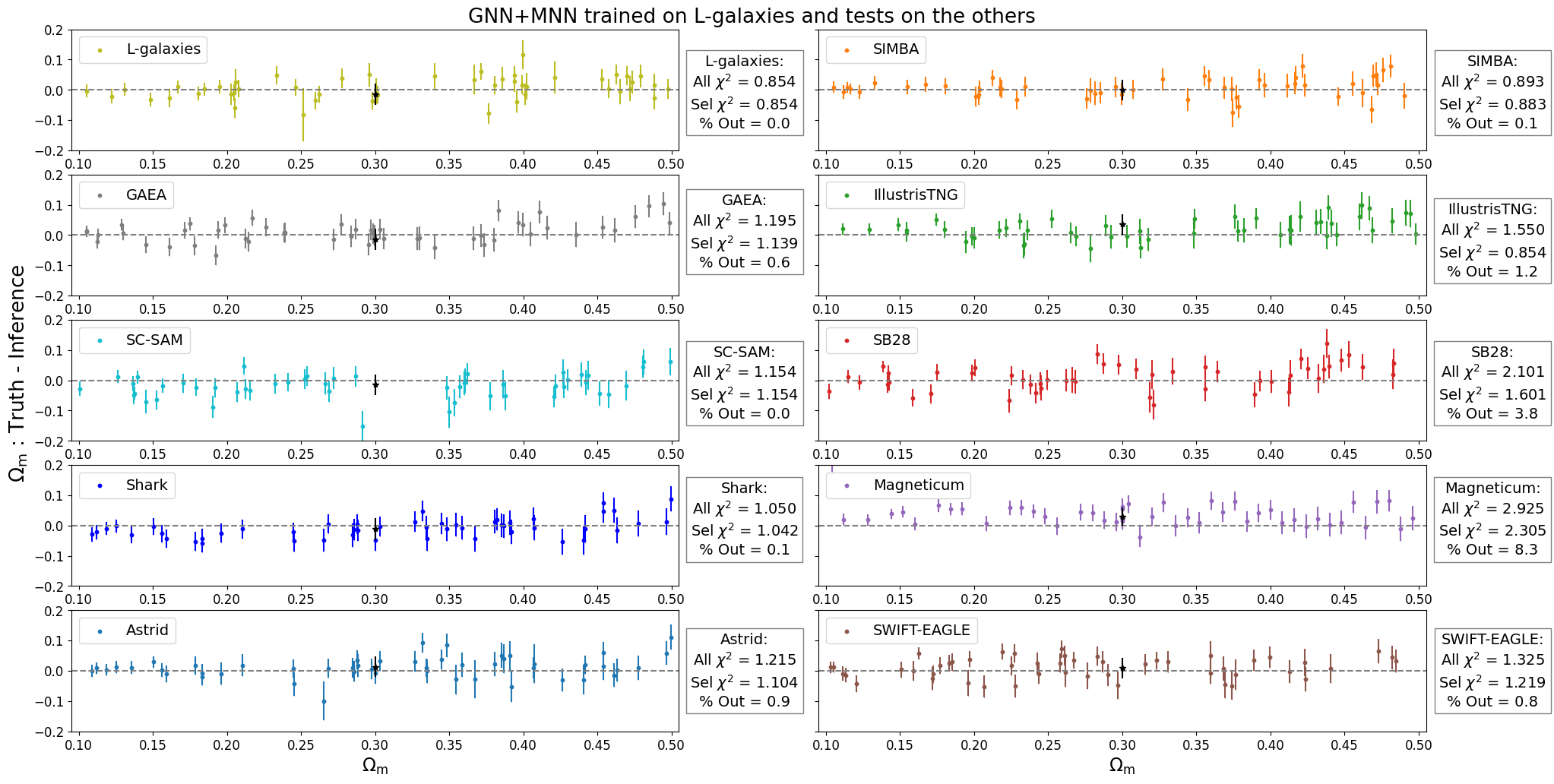}
 \caption{{\bf Truth - Inference values for $\mathbf{\Omega_m}$: GNN + MNN.} The model trained on L-Galaxies is evaluated on L-Galaxies, GAEA, SC-SAM, \texttt{Astrid}, {\sc SIMBA}, IllustrisTNG, SB28, Magneticum, and SWIFT-EAGLE catalogs. We plot the deviation of our predicted $\Omega_\textrm{m}$ from the true values (truth - $\mu$; see Equation \ref{eq:mu}), with error bars corresponding to $\sigma$ (see Equation \ref{eq:sigma}) predictions. Colored points represent the testing points (LH or SB sets), while the black points correspond to the average over the CV set inferences.
 We also present in the side legends the $\chi^2$ score (see Equation \ref{eq:chi2}) for all samples in the test sets and the selected samples, as well as the percentage of samples removed per test set. Overall, the model remains robust across all simulations considered, including both SAMs and hydrodynamical simulations.}
 \label{fig:main_result}
\end{figure}

%%%%%%%%%%%%%%%%%%%%%%%%%%%%%%%
\subsection{Training procedure and optimization choices}
\label{sec:train_optm}

We train our models on graphs constructed from galaxy catalogs of the LH sets from L-Galaxies. 
We initially split the $1,000$ LH simulations into training ($850$ simulations), validation ($100$ simulations), and testing ($50$ simulations) sets. 
For each simulation, we generate $10$ galaxy catalogs constructed by taking all galaxies with stellar masses larger than $1.3 R \times 10^8~M_\odot/h$, where $R$ is a random number uniformly distributed between $1$ and $2$. 
This strategy is made in order to marginalize over different minimum threshold values for stellar masses, as well to increase the number of catalogs used to train the model. 
The same trick was employed on \cite{deSanti2023-1}.
For each catalog, we produce a graph as outlined in Section \ref{sec:the_graph}.

We then train the model using the above architecture for $300$ epochs making use of the \textsc{Adam} optimizer \citep{Adam} to perform the gradient descent, and a batch size of $25$ samples.
The hyperparameter optimization (where we have used the learning rate, the weight decay, the linking radius, the number of message passing layers, and the number of hidden channels per layer of the MLPs) was carried out using the {\em Optuna} package \citep{optuna_2019} to perform a Bayesian optimization with Tree Parzen Estimator (TPE) \citep{Bergstra2011}. 
We made use of at least $100$ trials to perform this task and we directed the {\em Optuna} to minimize the validation loss, computed using an early-stopping scheme, in order to save only the model with the minimum validation error. The selected model was used for test subsequently. 
The best set of parameters are in Table \ref{tab:hyperparameters}.

We also trained the network on GAEA, but the resulting model was not robust, and therefore we do not present those results in detail.
Further discussion of this point can be found in Section \ref{sec:number}.

%%%%%%%%%%%%%%%%%%%%%%%%%%%%%%%%%%%%
\section{Results}
\label{sec:results}
%%%%%%%%%%%%%%%%%%%%%%%%%%%%%%%%%%%%

In this section, we present the main results obtained with the GNN-MNN model trained on L-Galaxies and tested across the different SAM prescriptions and hydrodynamical simulations.
Section \ref{sec:main_results} reports the model performance for both the cosmological and astrophysical parameter variations (LH and SB28 testing sets), as well as for the fiducial cosmology (CV set).
In Section \ref{sec:number}, we analyze the impact of the number of galaxies per catalog on the robustness of the model.

%%%%%%%%%%%%%%%%%%%%%%%%%%%%%%%%%%%%
\subsection{Field-Level Inference and Performance Metrics}
\label{sec:main_results}

\begin{wrapfigure}{l}{0.5\textwidth}
%\begin{figure}[h!]
    \centering
 \includegraphics[scale=0.45]{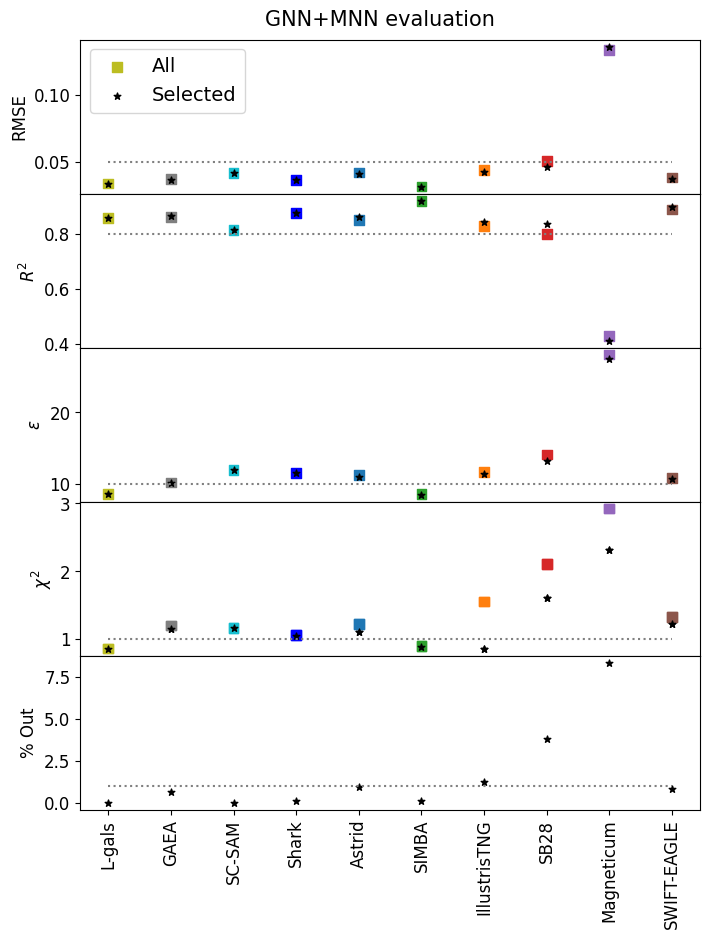}
    \caption{{\bf GNN+MNN: Model Evaluation.} We present RMSE (root mean square error), $R^2$ (coefficient of determination), $\epsilon$ (mean relative error), $\chi^2$ (reduced chi squared), and $\%$ Out (percentage of samples removed) for the test sets of L-Galaxies, GAEA, SC-SAM, {\sc SIMBA}, IllustrisTNG, SB28, Magneticum, and SWIFT-EAGLE indicated in the x-axis. Results are presented to all and selected samples according to $\chi^2$ values.}
    \label{fig:metrics_main}
%\end{figure}
\end{wrapfigure}

In Figure \ref{fig:main_result}, we present the difference between the inferred and true values of $\Omega_{\rm m}$ as a function of the true values.
The true values are represented accordingly to the first moment $\mu$ (see Equation \ref{eq:mu}) as well as their error bars correspond to the second moment $\sigma$ (see Equation \ref{eq:sigma}), predicted by the MNN (see more details in Section \ref{sec:free_likelihood_inference}).
The model was evaluated on the following galaxy formation models: L-Galaxies (testing set), GAEA, SC-SAM, {\sc Shark}, \texttt{Astrid}, {\sc SIMBA}, IllustrisTNG, SB28, Magneticum, and SWIFT-EAGLE.
Horizontal gray dashed lines indicate the ideal case of perfect predictions (i.e., zero residual).

For clarity and to avoid overcrowding the figure, we display only $50$ randomly chosen samples from each dataset, along with the average inference over the $27$ realizations of the CV set, represented by black star markers.
Additionally, we report the values of $\chi^2$ (see Equation \ref{eq:chi2}) in the figure legends, comparing the results for the entire test set (``all'') and for the subset obtained after excluding a fraction of catalogs with large residuals (``sel'').
This excluded fraction, also indicated in the legend, corresponds to catalogs with $\chi^2 > 10$ (see the definition of $\chi^2$ in Equation \ref{eq:chi2} as well as more information related to other evaluation metrics in the Appendix \ref{sec:metrics}). 
The same selection procedure was adopted in \cite{deSanti2023-2}, where predictions were filtered using the same $\chi^2$ threshold.
Even though this selection improves performance, the reason why the model performs poorly on these catalogs remains unclear and will require further investigation in future work.

The first test was performed using the model trained on L-Galaxies and evaluated on its corresponding L-Galaxies testing set.
Ultimately, the model provides accurate predictions for both the mean ($\mu$) and dispersion ($\sigma$) moments across the LH samples (covering the full $\Omega_{\rm m}$ range) and the CV realizations.
No catalogs needed to be excluded from the testing set, confirming the good performance of the model.

The second analysis evaluates the model's performance on the GAEA, SC-SAM, and {\sc Shark} galaxy catalogs.
In general, the predictions remain accurate across the entire range of $\Omega_{\rm m}$ values, including for the CV realizations.
A small fraction of catalogs ($0.6 \%$ for GAEA and $0.1\%$ for {\sc Shark}) were excluded for exceeding the imposed $\chi^2$ threshold.
However, these represent a negligible portion of the datasets, and the corresponding $\chi^2$ values for the selected predictions remain close to unity, confirming that the model extrapolates well across the different SAM prescriptions.

Finally, the third analysis evaluates the model on the hydrodynamical simulations.
The predictions exhibit small error bars and values mostly centered around zero, although the corresponding $\chi^2$ values are noticeably higher.
This effect is particularly evident for SB28 (all $\chi^2 = 2.10$) and Magneticum (all $\chi^2 = 2.93$).
Another key difference, when compared to the performance on the SAMs, lies in the fraction of catalogs removed due to the $\chi^2$ selection cut.
These fractions remain below or around $1 \%$ for \texttt{Astrid}, {\sc SIMBA}, IllustrisTNG, and SWIFT-EAGLE, but increase to $3.8\%$ and $8.3\%$ for SB28 and Magneticum, respectively.

In Figure \ref{fig:metrics_main}, we report several performance metrics: RMSE (root mean square error, measuring the typical prediction error; Equation \ref{eq:RMSE}),
$R^2$ (coefficient of determination, quantifying how much variance is explained by the model; Equation \ref{eq:R2}),
$\epsilon$ (mean relative error, expressing the average fractional deviation; Equation \ref{eq:rel_error}),
$\chi^2$ (reduced chi-squared, assessing consistency between predictions and uncertainties; Equation \ref{eq:chi2}),
and the percentage of catalogs excluded based on their $\chi^2$ values (\% Out).
Full definitions of these evaluation metrics are provided in Appendix \ref{sec:metrics}.
These metrics are shown for the model tested on L-Galaxies, GAEA, SC-SAM, {\sc Shark}, \texttt{Astrid}, {\sc SIMBA}, IllustrisTNG, SB28, Magneticum, and SWIFT-EAGLE.
For each case, we present results computed over both the entire testing set (``all'') and the selected subset of catalogs (``selected'').

Across all tests, the model exhibits consistently strong performance: RMSE values remain low (around $0.05$), $R^2$ values are close to $0.8$ (approaching unity), and the relative errors ($\epsilon$) are typically near $10 \%$.
The fraction of catalogs excluded due to high $\chi^2$ values is generally below $2.5 \%$ for most galaxy formation models.
A moderate decline in performance is observed for SB28 and Magneticum, which show higher RMSE, $\epsilon$, and $\chi^2$ values, along with a larger fraction of excluded catalogs (exceeding $7.5 \%$ for Magneticum).
Nevertheless, all metrics remain largely stable after applying the $\chi^2 < 10$ selection, indicating that outlier removal does not significantly alter the overall model performance.

The achieved precision of $\sim 10\%$ is consistent with expectations for field-level inference using galaxy phase-space information alone (e.g., \cite{pablo-galaxies-2022, deSanti2023-1}). In particular, the model trained and tested on L-Galaxies reaches an accuracy of $\sim 8.6\%$, improving over the $\sim 12\%$ obtained when training and testing on \texttt{Astrid} in \cite{deSanti2023-1}, likely due to the broader diversity of SAM-based catalogs. However, when evaluated on other SAMs and hydrodynamical simulations, the precision degrades to $\sim 12\%$, reflecting the challenges of extrapolating across different galaxy formation models. This behavior highlights an intrinsic trade-off between accuracy and robustness, and can be attributed to the limited information content of the input features (positions and line-of-sight velocities), the finite simulation volume, and degeneracies with astrophysical processes. Further improvements are expected from larger volumes, more diverse training datasets, and the inclusion of additional observables that preserve robustness.

While the best extrapolation performance is achieved across the different SAMs rather than when applying the model on the hydrodynamical simulations, the model still extrapolates well regardless of the merger tree similarities and differences among the datasets.
The model was trained on L-Galaxies, whose merger trees are based on the DM-only counterpart of \texttt{Astrid} (as is the case for GAEA, SAM-SC, and {\sc Shark}), and yet it not only generalizes well to \texttt{Astrid}, but also to {\sc SIMBA}, IllustrisTNG, Magneticum, and SWIFT-EAGLE.

%%%%%%%%%%%%%%%%%%%%%%%%%%%%%%%%%%%%
\subsection{The influence of the number of galaxies per catalog}
\label{sec:number}

\begin{figure}[h!]
 \centering
 \includegraphics[scale=0.39]{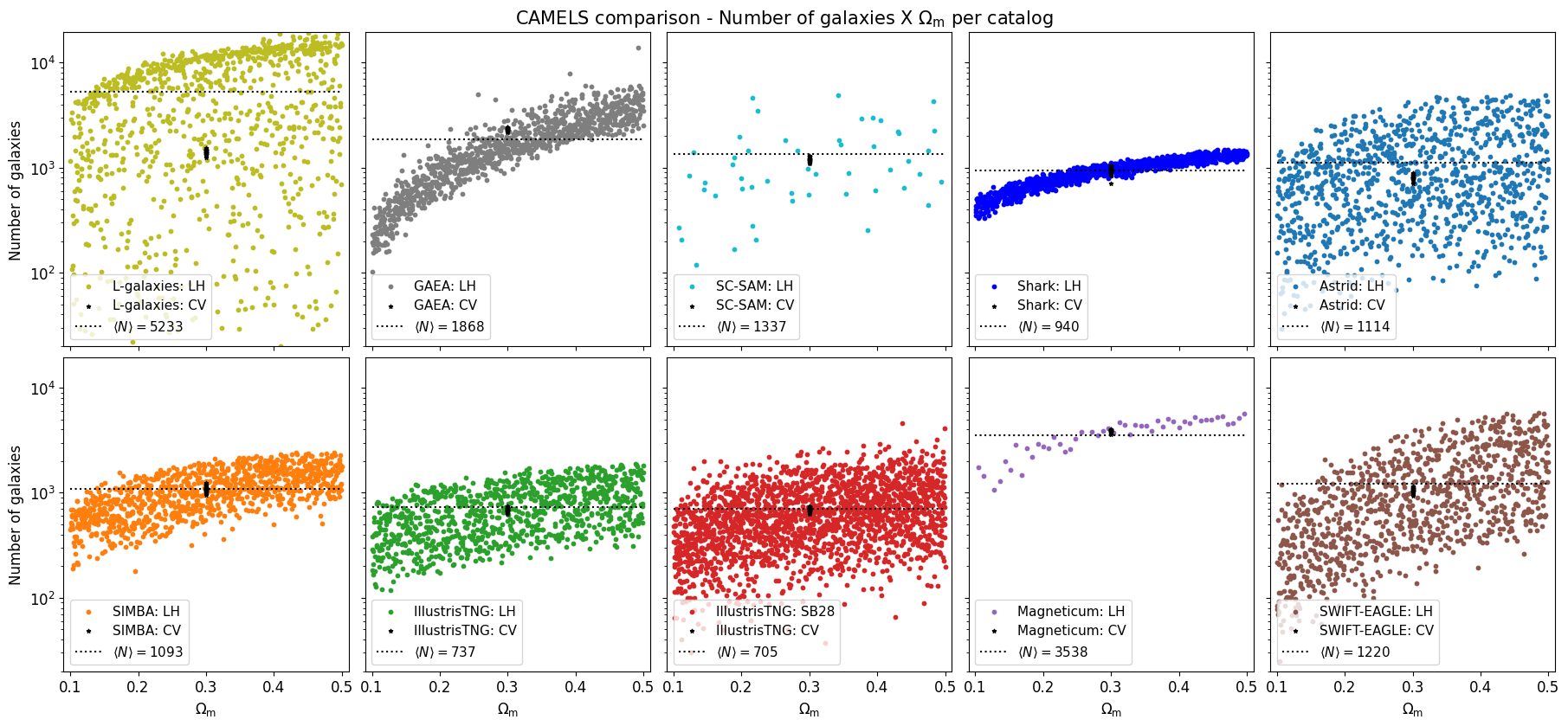}
 \caption{{\bf Comparison of the number of galaxies per $\Omega_m$ value.} We present the number of galaxies per $\Omega_m$ value for L-Galaxies, GAEA, SC-SAM, Shark, \texttt{Astrid}, IllustrisTNG, {\sc SIMBA}, SB28, Magneticum, and SWIFT-EAGLE catalogs. The horizontal dotted lines correspond to the mean number of galaxies per different model, while the black points represent the CV catalogs.}
 \label{fig:Nxomegam}
\end{figure}

One of the main findings in \cite{deSanti2023-1} was that the success of our previous model could be attributed to the use of a diverse training dataset, particularly in terms of the number of galaxies per catalog across different $\Omega_{\rm m}$ values.
A similar analysis is presented in this work, where we compare the number of galaxies per catalog as a function of $\Omega_{\rm m}$ in Figure \ref{fig:Nxomegam}.
In these plots, each point represents an individual catalog from the LH or SB samples (colored according to the corresponding simulation) and from the CV set (shown as black points).
Horizontal black dotted lines indicate the average number of galaxies for each galaxy formation model, with the corresponding values also reported in the legend.

As expected due to the success of the model trained on L-Galaxies, their catalogs exhibit a wide variation in the number of galaxies per realization, ranging from just one up to $19,308$. 
This spread is even broader than that found in \texttt{Astrid} and encompasses all other galaxy formation models.
Moreover, the average number of galaxies in L-Galaxies is the highest among all models ($\langle N \rangle = 5,233$), while the corresponding value for the CV set follows the general trend of about $1,000$ galaxies ($\langle N_{\mathrm{CV}} \rangle = 1,398$).
This extreme variation arises from the parameter choices adopted in the L-Galaxies LH samples used to generate their galaxy catalogs.

The other SAMs (GAEA, {\sc Shark}, and SC-SAM) do not exhibit as broad a variation in the number of galaxies, but they follow the same trend as the hydrodynamical simulations, with the number of galaxies increasing as $\Omega_m$ increases. 
Their average numbers of galaxies per catalog are $\langle N \rangle = 1,868$, $\langle N \rangle = 1,337$, and $\langle N \rangle = 940$, respectively for GAEA, {\sc Shark}, and SC-SAM, which are consistent with the average values found in the hydrodynamical simulations. 
Nevertheless, a significant variation in the number of galaxies is observed for SC-SAM, suggesting that variations in its parameters may lead to a pattern similar to that seen in L-Galaxies. However, additional realizations of the model are required to confirm this claim.

Finally, the superior diversity of L-Galaxies becomes clear when training the model on other SAMs.
Models trained on GAEA, for example, are significantly less robust than those trained on L-Galaxies.
This can be attributed to the fact that GAEA does not produce a sufficiently wide range of galaxy counts, reducing the diversity seen by the model and limiting its ability to generalize.
As a result, the model trained on GAEA struggles to predict $\Omega_{\rm m}$ for other simulations.

This behavior also provides insight into the impact of the training dataset choice.
It mirrors the findings of \cite{deSanti2023-1}, where models trained on IllustrisTNG or {\sc SIMBA} exhibited weaker generalization compared to those trained on the more diverse \texttt{Astrid} dataset.
Together, these results suggest that training on datasets with broader galaxy population diversity, such as L-Galaxies, is a key requirement for achieving robust extrapolation across simulations.

%Finally, the superior diversity of L-Galaxies becomes clear when training the model on other SAMs.
%Models trained on GAEA, for example, are significantly less robust than those trained on L-Galaxies.
%This can be attributed to the fact that GAEA does not produce a sufficiently wide range of galaxy counts, reducing the diversity seen by the model and limiting its ability to generalize.
%As a result, the model trained on GAEA struggles to predict $\Omega_{\rm m}$ for other simulations.
%This mirrors the findings of \cite{deSanti2023-1}, where models trained on IllustrisTNG or {\sc SIMBA} also exhibited weaker generalization compared to those trained on the more diverse \texttt{Astrid} dataset.

Another contributing factor may be differences in the galaxy properties produced by L-Galaxies compared to those found in the other SAMs.
For example, \cite{Ni2023} reported that \texttt{Astrid} exhibits larger variations in several galaxy properties across its LH parameter space-variations that may also be present in L-Galaxies but not in models such as GAEA.
Nevertheless, a direct comparison is still needed to identify which specific galaxy properties contribute to the success of the model trained on L-Galaxies, as well as how these properties differ across the other SAMs and the hydrodynamical simulations.

%%%%%%%%%%%%%%%%%%%%%%%%%%%%%%%%%%%%
\section{Discussion and Conclusions}
\label{sec:conclusions}
%%%%%%%%%%%%%%%%%%%%%%%%%%%%%%%%%%%%

Many different approaches have been, and continue to be, developed to construct comprehensive models of halo and galaxy formation and evolution, as well as to connect these models with observable properties \citep{Henriques2020, DeLucia2014, Perez2023, Lagos2024, Astrid2022, SIMBA2019, Pillepich2018, Hirschmann2014MNRAS, EAGLE2015}.
From hydrodynamical simulations to semi-analytic models (SAMs), subhalo abundance matching (SHAM), and halo occupation distribution (HOD) frameworks, all of these methods aim to reproduce the Universe across its diverse scales and properties \citep{Wechsler2018, Somerville2015}. 
However, it remains unclear which of these approaches best reproduces the relevant observables, and many of them, particularly hydrodynamical simulations, are prohibitively expensive to run at the scales needed for modern cosmological analyses.

At the same time, the pursuit of the most effective tools for generating mock catalogs and predicting cosmological parameters, with the highest possible accuracy and precision for current and upcoming observational surveys \citep{SKA1999, Laureijs2011, Ellis2012, Amendola2012, Benitez2014, Roman2015, DESI.DR2.DR2, Euclid2022-Tiago_Castro, JWST}, has never been more pressing.
While numerous models exist to describe galaxy formation, machine learning (ML) techniques are emerging as powerful tools that can assist the community in making more data-driven and informed choices, even when integrated with traditional analysis pipelines \citep{deSanti2022}.

In this context, graph neural networks (GNNs) emerge as powerful tools for analyzing galaxy catalogs because:
(1)	they are inherently designed to handle sparse and irregular data structures \citep{Gilmer2017, Battaglia2018, Bronstein2021};
(2) it is straightforward to construct GNN architectures that respect physical symmetries \citep{pablo-galaxies-2022}; 
and (3) they can extract information from galaxy catalogs without imposing a predefined scale cutoff. 
Furthermore, GNNs have already been employed to develop robust models across different hydrodynamical simulations \citep{deSanti2023-1, Roncoli2023} and DM-only simulations \citep{Shao2023-2, helen-halos-2022}, and have also proven effective for incorporating systematic effects \citep{deSanti2023-2}.

In this work, we trained a GNN coupled with a moment neural network (MNN) on thousands of catalogs generated with the fast semi-analytic model L-Galaxies, in order to infer $\Omega_{\rm m}$ at the field level.
More importantly, we assessed the robustness of the model by testing it on galaxy catalogs generated from simulations that employ completely different codes from those used for training.
Our main findings can be summarized as follows:

\begin{itemize}

\item The model trained on L-Galaxies, using only galaxy positions and velocities, is capable of inferring $\Omega_{\rm m}$ with an accuracy and precision of approximately $8.6\%$ when tested on L-Galaxies catalogs with varying cosmological and astrophysical parameters.

\item The robustness tests performed across different SAM prescriptions, specifically, GAEA, SC-SAM, and {\sc Shark}, show that the model's relative error remains below $12 \%$, even without excluding catalogs based on the $\chi^2$ threshold (see Section \ref{sec:main_results}). Moreover, more than $99.4 \%$ of catalogs are retained after applying the $\chi^2$ selection, indicating stable performance across different semi-analytic implementations.

\item The GNN model demonstrates remarkable extrapolation power when applied to hydrodynamical simulations. Its accuracy and precision remain below $12 \%$ without excluding any catalogs for \texttt{Astrid}, {\sc SIMBA}, IllustrisTNG, and SWIFT-EAGLE. Only SB28 and Magneticum exhibit higher relative errors ($14 \%$ and $28 \%$, respectively), reflecting differences in cosmological and astrophysical parameter variations compared to those used for training, as well as possible intrinsic discrepancies between the Magneticum and L-Galaxies models. In all cases, the fraction of catalogs exceeding the $\chi^2$ threshold remains below $8.3 \%$, underscoring the overall robustness of the approach.
\end{itemize}

Importantly, the GNN model presented here remains robust across different prescriptions for satellite galaxies.
Although the model was trained on L-Galaxies which, like GAEA, models satellite galaxies based on subhalo merger trees, it still achieves strong predictions for SC-SAM and {\sc Shark}, both of which model satellites using simpler NFW halo mass profiles, after the subhalos disappears from being tracked.
Moreover, the model performs well on the hydrodynamical simulations, where satellite and central galaxies are identified directly through numerical solutions of the underlying physics, rather than analytic prescriptions.
This demonstrates that the model is not sensitive to the specific theoretical choices underlying satellite modeling, further reinforcing its robustness.

A meaningful point of comparison is between the GNN model presented here and our previous model in \cite{deSanti2023-1}.
The model trained on L-Galaxies and tested on L-Galaxies achieves an accuracy of approximately $8.6\%$, representing a notable improvement over the model trained and tested on \texttt{Astrid} in \cite{deSanti2023-1}, which reached an accuracy of about $12\%$.
For hydrodynamical simulations such as SIMBA and IllustrisTNG, the accuracies remain broadly comparable between the two works (around $10\%$).
However, the previous model performed better on SB28 and Magneticum (below $13\%$), highlighting intrinsic differences between SAM- and hydro-based datasets and the challenges posed by their distinct galaxy populations.
We note that the model in \cite{deSanti2023-1} was designed specifically for hydrodynamical simulations and therefore was not intended to be applied to SAMs.
The present work extends this line of investigation by evaluating robustness across both SAMs and hydrodynamical simulations.

Additionally, in Appendix \ref{sec:NF}, we present a slight adaptation of the model, in which the GNN is coupled to a normalizing flows (NF) instead of the usual MNN.
Although this adapted model also exhibits signs of robustness, we emphasize that these results are not among the main findings of this paper.
The key conclusion from this exercise is that a larger number of simulations will be required to accurately predict full posterior distributions when employing generative approaches such as NFs.
This insight will guide future work, where new simulation models will be incorporated to further explore and improve generative cosmological inference frameworks.

%The potential application of this framework to real observational data depends on an inherent limitation of the current methodology, one that remains closely tied to the question of robustness.
%Specifically, a GNN can reliably extrapolate its predictions only when applied to data that are compatible with the domain on which it was trained.
%\natali{In this work we gave an important step, showing that these ML methods can be applied to SAMs, that are way feasible to run, if compared to the hydrodynamical simulations. 
%More than that, the machinery we presented showcase SAMs as  largely independent of their specific physical prescriptions, reinforcing their potential as realistic mock catalogs for cosmological parameter inference.}
%Although, a successful application to real data requires that the CAMELS (or any other) suite of simulations adequately captures the key characteristics of our observable Universe.
%For instance, many observational mock catalogs are constructed using light-cone geometries and HOD-based methods, which differ from the setup used in this study.

The potential application of this framework to real observational data depends on an inherent limitation of the methodology, one that remains closely tied to the question of robustness.
A GNN model can reliably extrapolate its predictions only when applied to data that are compatible with the domain on which it was trained.
In this work, we take an important step forward by demonstrating that these ML methods can be successfully applied to SAMs, which are far more computationally feasible to generate than hydrodynamical simulations.
Moreover, our results show that SAMs exhibit a remarkable level of internal consistency, being largely independent of their specific physical prescriptions.
This strongly reinforces their potential as realistic and efficient way to produce catalogs for field-level cosmological inference.

Nevertheless, applying these GNN models to real data will likely require larger simulation suites, whether CAMELS or others, that incorporate key observational effects such as light-cone geometries and survey selection functions.
These ingredients are typically implemented through HOD-based mock constructions, which differ from the idealized setups used in this study.
Integrating such observational features is therefore a crucial next step toward developing robust, SAM-based ML inference pipelines for real survey data.

In future work, we plan to address these challenges by extending our analysis to additional approximate methods and by incorporating redshift-space distortions and light-cone maps.
Our goal is to assess whether the GNN model can remain robust, accurate, and precise enough to infer not only $\Omega_{\rm m}$ but also other cosmological parameters from more realistic galaxy catalogs.

%% IMPORTANT! The old "\acknowledgment" command has be depreciated. It was
%% not robust enough to handle our new dual anonymous review requirements and
%% thus been replaced with the acknowledgment environment. If you try to 
%% compile with \acknowledgment you will get an error print to the screen
%% and in the compiled pdf.
%% 
%% Also note that the akcnowlodgment environment does not support long amounts of text. If you have a lot of people and institutions to acknowledge, do not use this command. Instead, create a new \section{Acknowledgments}.

\section*{Acknowledgments}
NSMS acknowledges partial support from the Simons Foundation, NSF CDSE grant AST-2408026 and the NASA TCAN grant 80NSSC24K0101.
FF acknowledges support from the Next Generation European
Union PRIN 2022 ``20225E4SY5 - From ProtoClusters to Clusters in one Gyr''. PA-A acknowledges support from the Independent Research Fund Denmark (DFF) under grant 4251-00086B. 
VGP has been supported by the Atracci\'{o}n de Talento Contract no. 2023-5A/TIC-28943 granted by the Comunidad de Madrid in Spain, and the national grants CNS2024-154242, PID2024-159420NB-C43, and PID2021-122603NB-C21. 

The run and analysis for {\sc Shark} has been carried out in the computing cluster at UAM (\textsc{taurus}).
L-Galaxies, GAEA, and SC-SAM have been run on the Simons Foundation, Flatiron Institute, Center of Computational Astrophysics computing cluster.
The training of the GNNs has been carried out using  graphics processing units (GPUs) from Simons Foundation, Flatiron Institute, Center of Computational Astrophysics.
The Flatiron Institute is supported by the Simons Foundation.
OpenAI's language model – ChatGPT was used for some language editing during the draft stage of the preparation of this manuscript.

%% To help institutions obtain information on the effectiveness of their 
%% telescopes the AAS Journals has created a group of keywords for telescope 
%% facilities.
%
%% Following the acknowledgments section, use the following syntax and the
%% \facility{} or \facilities{} macros to list the keywords of facilities used 
%% in the research for the paper.  Each keyword is check against the master 
%% list during copy editing.  Individual instruments can be provided in 
%% parentheses, after the keyword, but they are not verified.

\vspace{5mm}
\facilities{Flatiron Institute and UAM ({\sc taurus}) computing clusters.}

%% Similar to \facility{}, there is the optional \software command to allow 
%% authors a place to specify which programs were used during the creation of 
%% the manuscript. Authors should list each code and include either a
%% citation or url to the code inside ()s when available.

\software{Pytorch \citep{pytorch-geometric},  
          {\em Optuna} \citep{optuna_2019}, 
          \textsc{CosmoGraphNet} \citep{pablo-galaxies-2022},
          {\em Scikit-learn} \citep{scikit-learn},
          Pyro \citep{bingham2019pyro},
          TARP \citep{lemos2023}.
          }

%% Appendix material should be preceded with a single \appendix command.
%% There should be a \section command for each appendix. Mark appendix
%% subsections with the same markup you use in the main body of the paper.

%% Each Appendix (indicated with \section) will be lettered A, B, C, etc.
%% The equation counter will reset when it encounters the \appendix
%% command and will number appendix equations (A1), (A2), etc. The
%% Figure and Table counter will not reset.

\appendix

%%%%%%%%%%%%%%%%%%%%%%%%%%%%%%%
\section{SAMs specifications}
\label{sec:SAM+}
%%%%%%%%%%%%%%%%%%%%%%%%%%%%%%%

In this appendix we present the specifications related to the SAMs utilized to train, validate and test the ML models, which are first presented in Section \ref{sec:simulations}. 
The parameters being varied as well as their ranges and fiducial values are described in Tables \ref{tab:LGalaxies_parameters}, \ref{tab:gaea_parameters}, \ref{tab:S-SAM_parameters}, and \ref{tab:shark_parameters}, respectively for L-Galaxies, GAEA, SC-SAM, and {\sc Shark}.

\begin{table*}[h!]
  \caption{Summary of the parameters sampled in L-Galaxies. The equations related to the free parameters are presented in the Supplementary Material of \citet{Henriques2020}. \label{tab:LGalaxies_parameters}}
    \begin{center}
        \begin{tabular}{lccc}
        \hline\hline
        \textbf{Parameter} & \textbf{Fiducial Value} &\textbf{Range} & \textbf{Reference}  \\
        \hline\hline
        {\bf Star Formation} & & &\\
        $\alpha_{\rm H_{2}}$ & $0.06$ &$10^{-12} - 10^{7}$ &  Eq.(S16) \\
        $\alpha_{\rm SF, burst}$ & $0.5$ &$10^{-4}-1.0$ & Eq.(S37) \\
        $\beta_{\rm SF, burst}$ & $0.38$ &$10^{-4}- 1.0$ & Eq.(S37) \\
        {\bf AGN feedback \& BH growth} & & & \\
        $k_{\rm AGN}\ [M_{\odot} \ {\rm yr}^{-1}]$ & $2.5 \cdot 10^{-3}$ & $10^{-8}- 1 $  & Eq.(S28) \\
        $f_{\rm BH}$ & $0.066$ &$0.01-1$ & Eq. (S27) \\  
        $V_{\rm BH} \ [{\rm km \ s}^{-1}]$ & $700$&$1- 3 \cdot 10^{3}$ & Eq.(S27) \\
        {\bf Stellar Feedback} & & &\\
        $\epsilon_{\rm reheat}$ & $5.6$& $10^{-4}-10$  & Eq.(S22) \\
        $V_{\rm reheat}\ [{\rm km \ s}^{-1}]$ & $110$ & $1- 10^{3}$ & Eq.(S22) \\
        $\beta_{\rm reheat}$ & $2.9$ &$10^{-4}-5$ & Eq.(S22) \\
        $\eta_{\rm eject}$ & $5.5$ &$10^{-4}- 10$ & Eq.(S20) \\
        $V_{\rm eject}\ [{\rm km \ s}^{-1}]$ & $220 $ & $1- 10^{3} $ & Eq.(S20) \\
        $\beta_{\rm eject}$ & $2.0$ & $10^{-4}- 5$ & Eq.(S20) \\
        {\bf Reincorporation} & & & \\
        $\gamma_{\rm reinc}\ [{\rm yr}^{-1}]$ & $1.2 \cdot 10^{10}$ & $10^{-7}-10^{15}$ & Eq.(S25) \\
        {\bf Tidal Disruption} & & &\\
        $\alpha_{\rm friction}$ & $1.8$ & $0.01-10$ & Eq.(S26) \\
        \hline
        \end{tabular}
    \end{center}
\end{table*}

\begin{table*}[h!]
  \caption{\label{tab:gaea_parameters} Summary of the parameters sampled in GAEA.}
	\begin{center}
		\begin{tabular}{cccc}
		\hline\hline
		\textbf{Parameter} & \textbf{Fiducial value} & \textbf{Range} & \textbf{Reference}  \\
		\hline\hline
		{\bf Efficiency of supernovae} & & & \\
		$\epsilon_{\rm reheat}$ & $0.028$ & $0-1$ & Tab.(1) in \cite{Hirschmann2016} \\
		$\epsilon_{\rm eject}$ & $0.10$ & $0-1$ & Tab.(1) in \cite{Hirschmann2016} \\
		{\bf AGN feedback} & \\
		$k_{\rm radio}$ & $1.36$ & $0-2 \cdot 10^{-4}$ & Eq.(2) in \cite{Fontanot2020}\\
		$\epsilon_{\rm qw}$ & $4.86$ & $0.5-10^3$ & Eq.(15) in \cite{Fontanot2020}\\
        \hline
		\end{tabular}
	\end{center}
\end{table*}

\begin{table*}[h!]
  \caption{\label{tab:S-SAM_parameters} Summary of the parameters sampled in SC-SAM. More details can be found in \textsection 2.3 \cite{Perez2023}.}
	\begin{center}
		\begin{tabular}{cccc}
		\hline\hline
		\textbf{Parameter} & \textbf{Fiducial value} & \textbf{Range}  & \textbf{Reference} \\
		\hline\hline
		{\bf Supernovae feedback} & & & \\
		$A_{\textrm{SN1}}$ & $1$ & $[0.25, 4]$ & Eq. (2) in \citet{Somerville_sam2015}\\
		$A_{\textrm{SN2}}$ & $0$ & $[- 2, 2]$ & Eq. (2) in \citet{Somerville_sam2015}\\
		{\bf AGN feedback} & & \\
		$A_{\textrm{AGN}}$ & $1$ & $[0.25, 4]$& Eq. (20) in \citet{Somerville2008}\\
        \hline
		\end{tabular}
	\end{center}
\end{table*}

\begin{table*}[h!]
 \caption{\label{tab:shark_parameters} Summary of the parameters sampled in {\sc Shark}.}
	\begin{center}
		\begin{tabular}{cccc}
		\hline\hline
		\textbf{Parameter} & \textbf{Fiducial value} & \textbf{Range} & \textbf{Reference}  \\
		\hline\hline
		{\bf Stellar Feedback} & & & \\
		$v_{\rm hot}$ &  $80\ \rm{km.s^{-1}}$ & $10^{-3}-500\ \rm{km.s^{-1}}$ & Eq. (25)-(28) in \cite{Lagos2018} \\
		$\beta$ & $3.79$ & $5\times10^{-3}-5$ & Eq. (25)-(28) in \cite{Lagos2018} \\
		$\beta_{\rm min}$ & $0.104$ & $0.01-1$ & Appendix A2 in \cite{Lagos2023} \\
		{\bf Star Formation} & \\
		$\nu_{\rm SF}$ &  $1.7\ \rm{Gyr^{-1}}$ & $0.01-10\ \rm{Gyr^{-1}}$ & Eq. (7) in \cite{Lagos2018} \\
		{\bf Reincorporation} & & & \\
		$\tau_{\rm reinc}$ & $21.53\ \rm{Gyr}$ & $10^{-3}-10^{3}\ \rm{Gyr}$ & Eq. (30) in \cite{Lagos2018} \\
		$M_{\rm norm}$ & $1.183\times10^{11}\ \rm{M_{\odot}}$ & $10^{9}-10^{12}\ \rm{M_{\odot}}$ & Eq. (30) in \cite{Lagos2018} \\
		$\gamma$ & $-2.339$ & $-3-0$ & Eq. (30) in \cite{Lagos2018} \\
		{\bf AGN feedback} \& BH growth & & & \\
		$f_{\rm smbh}$ & $0.008$ & $10^{-5}-100$ &  Eq. (37) in \cite{Lagos2018}\\
		$\rm{e_{sb}}$ & $20$ & $0.5-50$ & Section 4.4.10 in \cite{Lagos2018} \\
		$\eta$ &  $4$ & $1-10$ & Section 4.4.10 in \cite{Lagos2018} \\
		$\kappa$ & $10.31$ & $10^{-5}-100$ & Eq. (40) in \cite{Lagos2018} \\
		$\kappa_{\rm radio}$ & $0.023$ & $0.1-3$ & Eq. (32) in \cite{Lagos2023} \\
		$\epsilon_{\rm QSO}$ & $10$ & $0.1-100$ & Eq. (37) in \cite{Lagos2023} \\
		{\bf Tidal and Ram-pressure stripping} & & & \\
		$\alpha_{\rm RPS}$ &  $1$ & $0.1-10$ & Eq. (52) in \cite{Lagos2023} \\
		minimum\_halo\_mass\_fraction &  $0.01$ & $10^{-5}-0.1$ & Section 3.5 in \cite{Lagos2023} \\
		{\bf Disc instabilities} & \\
		$\epsilon_{\rm disc}$ &  $0$ & $0-50$ & Section 4.4.8 in \cite{Lagos2018} \\
        \hline
		\end{tabular}
	\end{center}
\end{table*}

%%%%%%%%%%%%%%%%%%%%%%%%%%%%%%%
\section{ML model details}
\label{sec:MLdetails}
%%%%%%%%%%%%%%%%%%%%%%%%%%%%%%%

In this appendix, we elaborate on the details of the graphs (see Section \ref{sec:graph_details}), of the employed architecture (Sections \ref{sec:architecture_details} and \ref{sec:free_likelihood_inference}), and present the definitions of the metrics (Section \ref{sec:metrics}) used to evaluate the ML models presented in this work through all the Sections \ref{sec:methodology} and \ref{sec:results}.

\subsection{Graph details}
\label{sec:graph_details}

As described in Section~\ref{sec:the_graph}, the graphs are constructed following the methodology of \cite{deSanti2023-1}, where galaxies are represented as nodes carrying information about their peculiar velocities, $v_z$.
This choice is primarily motivated by the analogy between $v_z$ and the line-of-sight velocity measured in observations.
The velocity information is incorporated as node attributes according to:
\begin{equation}
   v_z \rightarrow \mathrm{sign} (v_z) \cdot \log_{10} \left[ 1 + \mathrm{abs} (v_z) \right] . \label{eq:norm_vz}  
\end{equation}
We use the galaxy positions to establish the connections between galaxies, thereby defining the edges of the graphs.
In addition, the positions are used to compute the edge features, which are defined as:
\begin{equation}
 \mathbf{e}_{i j} = \left[ \frac{|\mathbf{d}_{i j}|}{r_{link}}, \alpha_{i j}, \beta_{i j} \right] \, ,
\end{equation}
where:
\begin{align}
 \mathbf{d}_{i j} & = \left[ \mathbf{r}_{i} - \mathbf{r}_{j} \right]\\
 \pmb{\delta}_{i} & = \mathbf{r}_{i} - \mathbf{c}\\
 \alpha_{i j} & = \frac{ \pmb{\delta}_{i}}{ |\pmb{\delta}_{i}| } \cdot \frac{ \pmb{\delta}_{j}}{ |\pmb{\delta}_{j}| }\\   
 \beta_{i j} & = \frac{ {\pmb \delta}_{i}}{ |{\pmb\delta}_{i}| } \cdot \frac{ \mathbf{d}_{i j}}{ |\mathbf{d}_{i j}| } \, ,
\end{align}
with $\mathbf{r}_{i}$ representing the position of a galaxy $i$ and $\mathbf{c}=\sum_i^N \mathbf{r}_i/N$ being the centroid.
Here, the {\em distance} $\mathbf{d}_{i j}$ is the difference of two galaxy ($i$ and $j$) positions, the {\em difference vector} $\pmb{\delta}_{i}$ denotes the position of a galaxy $i$ with respect to the centroid, $\alpha_{i j}$ is the (cosine of) the angle between the difference vectors of two galaxies,  while $\beta_{i j}$ represents the angle between the difference vector of a galaxy $i$ and its distance to another galaxy $j$. 
We account for periodic boundary conditions when computing both distances and angles. 

Every graph is characterized by a label that we aim at inferring ($\Omega_{\rm m}$). 
We normalize it using
\begin{equation}
 \Omega_m \rightarrow \frac{ \left( \Omega_m - \Omega_m^{\rm min} \right) }{ \left( \Omega_m^{\rm max} - \Omega_m^{\rm min} \right) } ,   
\end{equation}
where $\Omega_m^{\rm min}$ and $\Omega_m^{\rm max}$ represent the minimum and the maximum values.

\subsection{Architecture Details}
\label{sec:architecture_details}

We have used a message passing scheme cited in Section \ref{sec:architecture}, where each message passing layer updates the node and edge features, taking as input the graph and delivering as output its updated version. 
We do not employ a model to update the global attribute. 
The global attribute was used just in order to update the node information (see Equation \ref{eq:node_model}). 
The node and edge features at layer $\ell + 1$ are found from the node and edge features at layer $\ell$ as:

\begin{itemize}

 \item {\bf Edge model:}
  \begin{equation}
   \mathbf{e}_{i j}^{(\ell + 1)} = \mathcal{E}^{(\ell + 1)} \left( \left[ \mathbf{n}_{i}^{(\ell)}, 
   \mathbf{n}_{j}^{(\ell)}, \mathbf{e}_{i j}^{(\ell)} \right]  \right) ,
   \label{eq:edge_model}
  \end{equation}
  where $\mathcal{E}^{(\ell + 1)}$ represents a MLP;
 
 \item {\bf Node model:}
  \begin{equation}
   \mathbf{n}_{i}^{(\ell + 1)} = \mathcal{N}^{(\ell + 1)} \left( \left[ \mathbf{n}_{i}^{(\ell)}, 
   \bigoplus_{j \in \mathfrak{N}_i} \mathbf{e}_{i j}^{(\ell + 1)}, \mathbf{g} \right]  \right) ,
   \label{eq:node_model}
  \end{equation}
  where $\mathfrak{N}_i$ represents all neighbors of node $i$, $\mathcal{N}^{(\ell + 1)}$ is a MLP, and $\oplus$ is a multi-pooling operation responsible to concatenate several permutation invariant operations:
  \begin{equation}
   \bigoplus_{j \in \mathfrak{N}_i} \mathbf{e}_{i j}^{(\ell + 1)} = \left[ 
   \max_{j \in \mathfrak{N}_i} \mathbf{e}_{i j}^{(\ell + 1)},
   \sum_{j \in \mathfrak{N}_i} \mathbf{e}_{i j}^{(\ell + 1)},
   \frac{ \sum_{j \in \mathfrak{N}_i} \mathbf{e}_{i j}^{(\ell + 1)} }{ \sum_{j \in \mathfrak{N}_i} }
   \right] .  \label{eq:multi-pooling}
  \end{equation}
\end{itemize}
We also we made use of residual layers in the intermediate layers \citep{pablo-galaxies-2022}.

Once the graph has been updated using the $N$ message passing layers, we collapse it into a 1-dimensional feature vector using 
\begin{equation}
  \mathbf{y} = \mathcal{F} \left( \left[ \bigoplus_{i \in \mathcal{G}} \mathbf{n}_i^N, \mathbf{g}
  \right] \right) , \label{eq:last_layer}
\end{equation}
where $\mathcal{F}$ is the last MLP, $\oplus_{i \in \mathfrak{F}}$ the last multi-pooling operation (done exactly according to Equation \ref{eq:multi-pooling}, but operating over all nodes in the graph $\mathcal{G}$), and $\mathbf{y}$ represents the target of the GNN ($\Omega_{\rm m}$).

All the MLP are constructed by a series of fully connected layers with ReLU activation 
function (except for the last layer, which does not employ an activation function). 

\subsection{Moment Neural Networks and the loss function}
\label{sec:free_likelihood_inference}

The model is trained to infer the value of $\Omega_{\rm m}$ by predicting the marginal posterior mean $\mu_i$ and standard deviation $\sigma_i$ without making any assumption about the form of the posterior, 
i.e.
\begin{equation}
 \mathbf{y}_i(\mathcal{G}) = [\mu_i(\mathcal{G}), \sigma_i(\mathcal{G})] ,
\end{equation}
where
\begin{align}
 \mu_i ( \mathcal{G} ) & = \int_{ \Omega_m^i } d \Omega_m^i ~ \Omega_m^i ~ p (\Omega_m^i | \mathcal{G} ) \label{eq:mu}\\
 \sigma^2_i ( \mathcal{G} ) & = \int_{ \Omega_m^i } d \Omega_m^i ~ ( \Omega_m^i - \mu_i)^2 ~ p (\Omega_m^i | \mathcal{G} ) \, . \label{eq:sigma}
\end{align}
$\mathcal{G}$ represents the input graph and $p (\Omega_m^i | \mathcal{G} )$ is the marginal
posterior, taken according to
\begin{equation}
 p (\Omega_m^i | \mathcal{G} ) = \int_{ \Omega_m^i } d \Omega_m^1 d \Omega_m^2 \dots d \Omega_m^n  
 ~ p (\Omega_m^1, \Omega_m^2, \dots, \Omega_m^n | \mathcal{G} ) .
\end{equation}
In order to achieve this, we made use of a specific loss function following \cite{Jeffrey2020}:
\begin{equation}
 \mathcal{L} = 
 \log \left[ \sum_{ j \in \mathrm{batch} } \left( \Omega_m^j - \mu_{j} \right)^2 \right] + 
 \log \left\{ \sum_{ j \in \mathrm{batch} } \left[ 
 \left( \Omega_m^j - \mu_{j} \right)^2 - \sigma_{j}^2 \right]^2 \right\} , \label{eq:loss}
\end{equation}
where $j$ represents the samples in a given batch. 

\subsection{Evaluation metrics}
\label{sec:metrics}

We quantify the accuracy and precision of the model using different metrics that we describe below. 
These metrics have been used mainly in Section \ref{sec:main_results} and in the Appendix \ref{sec:NF}.
We consider the true value of the parameter in question for graph $i$ as $\Omega_m^i$, while we denote as $\mu_i$ and $\sigma_i$ the prediction of the network for the posterior mean and standard deviation, respectively.

\begin{itemize}

 \item {\bf Root Mean Squared Error (RMSE):}
  \begin{equation}
   \text{RMSE} = \sqrt{ \frac{1}{N} \sum_{i = 1}^N 
   \left( \Omega_m^i - \mu_i \right)^2 } . \label{eq:RMSE}
  \end{equation}
Low values of the RMSE indicate the model is precise.

 \item {\bf Coefficient of determination:}
  \begin{equation}
    R^2 = 1 - \frac{ \sum_{i = 1}^N \left( \Omega_m^i - \mu_i \right)^2 }{ 
    \sum_{i = 1}^N \left( \Omega_m^i - \bar{\Omega}_m^i \right)^2 } , \label{eq:R2} 
  \end{equation}
  where $\bar{\Omega}_m^i = \frac{1}{N} \sum_{i = 1}^N \Omega_m^i$. Values close to $1$ indicate the model is accurate, while negative values and values close to zero do not.

 \item {\bf Pearson Correlation Coefficient (PCC):}
  \begin{equation}
   \label{eq:PCC}
   \rm{PCC} = \frac{\rm{cov} \left( \Omega_m, \mu \right)}
       {\sigma_{\Omega_m} \sigma_\mu} .
  \end{equation}
  This statistic measures the positive/negative linear relationship between truth values 
  and inferences: good values are close to $\pm 1$, and worse closer to $0$. 

 \item {\bf Bias:}
  \begin{equation}
   b = \frac{1}{N} \sum_{i = 1}^N \left( \Omega_m^i - \mu_i \right) . \label{eq:bias}
  \end{equation}
This statistic quantifies how much the inferences are ``biased'' with respect to the truth values; unbiased values are close to $0$.
 
 \item {\bf Mean relative error:}
  \begin{equation}
   \epsilon = \frac{1}{N} \sum_{i = 1}^N \frac{ | \Omega_m^i - \mu_i | }{ \mu_i } . \label{eq:rel_error}
  \end{equation}
Low values of this statistic indicate the model is precise.
  
 \item {\bf Reduced chi squared:}
  \begin{equation}
   \chi^2 = \frac{1}{N} \sum_{i = 1}^N \left(\frac{\Omega_m^i - \mu_i }{\sigma_i} \right)^2  .  \label{eq:chi2}
  \end{equation}
  This statistic quantifies the accuracy of the estimated errors.
  Values of $\chi^2$ close to $1$ indicate the magnitude of the errors (posterior standard deviation in our case) is properly inferred, while values larger/smaller than $1$ indicate the model is under/over predicting the errors.
  This was the statistic chosen for selecting good and bad predictions (see Section \ref{sec:results}).
  
\end{itemize}

We make use of these statistics to quantify the accuracy, precision, and bias of the model on each different test sets (for L-Galaxies, GAEA, SC-SAM, {\sc Shark}, \texttt{Astrid}, {\sc SIMBA}, IllustrisTNG, SB28, Magneticum, and SWIFT-EAGLE). 
We have utilized {\em Scikit-learn} \citep{scikit-learn} for computing them.

\begin{figure}[h!]
 \centering
 \includegraphics[scale=0.34]{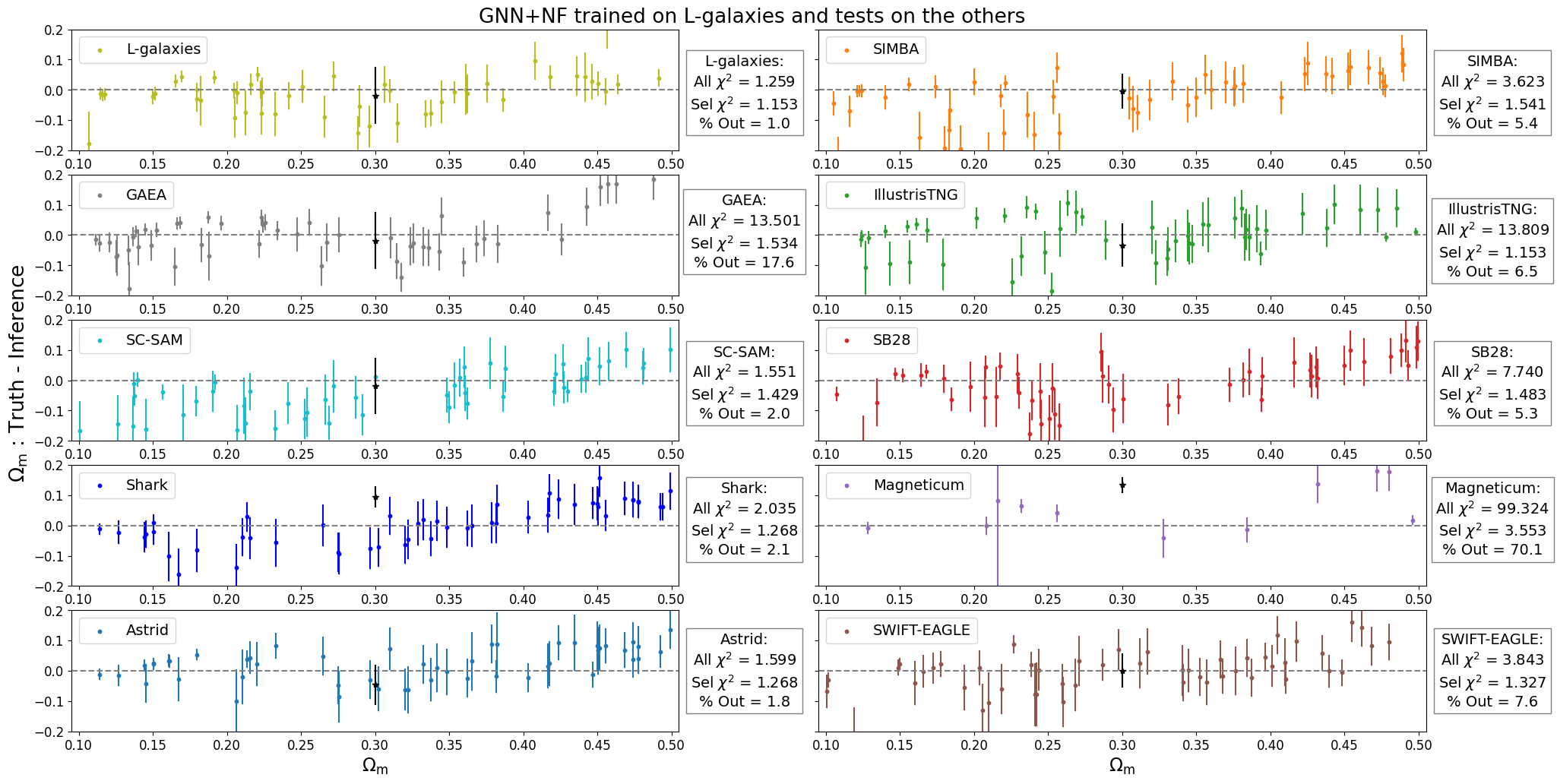}
 \caption{{\bf Truth - Inference values for $\Omega_m$: GNN + NF.} The model trained on L-Galaxies is evaluated on L-Galaxies, GAEA, SC-SAM, \texttt{Astrid}, {\sc SIMBA}, IllustrisTNG, SB28, Magneticum, and SWIFT-EAGLE catalogs. We plot the deviation of our predicted $\Omega_\textrm{m}$ from the true values (truth - $\mu$; see Equation \ref{eq:mu}), with error bars corresponding to $\sigma$ (see Equation \ref{eq:sigma}) predictions. Colored points represent the testing points (LH or SB sets), while the black points correspond to the average over the CV set inferences.
 We also present in the side legends the $\chi^2$ score (see Equation \ref{eq:chi2}) for all samples in the test sets and the selected samples, as well as the percentage of samples removed per test set. By contrast, this model does not achieve the same level of performance as the primary GNN-MNN model, showing reduced accuracy and less robust extrapolation across the different simulations.}
 \label{fig:result-GNN+NF}
\end{figure}

%%%%%%%%%%%%%%%%%%%%%%%%%%%%%%%
\section{GNN+NF}
\label{sec:NF}
%%%%%%%%%%%%%%%%%%%%%%%%%%%%%%%

Throughout this paper, we have presented predictions for $\Omega_{\rm m}$ using a GNN coupled with a MNN.
In this framework, the model predicts values that effectively touch the posterior distribution by estimating its first $\mu$ and second $\sigma$ moments (see Equations \ref{eq:mu} and \ref{eq:sigma}).
In this appendix, we extend the approach by coupling the GNN with a generative model based on NFs, enabling the network to infer the full posterior distribution of $\Omega_{\rm m}$ directly from the data.

The general idea of this method is to, starting with a random variable $\mathbf{x}$ with simple distribution $p (\mathbf{x})$, perform a transformation $f (\mathbf{x})$ or a series of transformations $f_{\phi} (\mathbf{x}) = f_1 \circ \dots \circ f_k (\mathbf{x})$ to learn a complex desired distribution for variable $\mathbf{z}$.
Then, the flow represents this bijective map, which needs to be invertible and differentiable, and the two distributions are connected via
\begin{equation}
   p (\mathbf{z}) = p (\mathbf{x}) \left| \det \left( \frac{\partial f^{- 1} (\mathbf{z})}{\partial \mathbf{x}} \right) \right| .
\end{equation}

\begin{wrapfigure}{l}{0.5\textwidth}
% \begin{figure}[h!]
    \centering
 \includegraphics[scale=0.45]{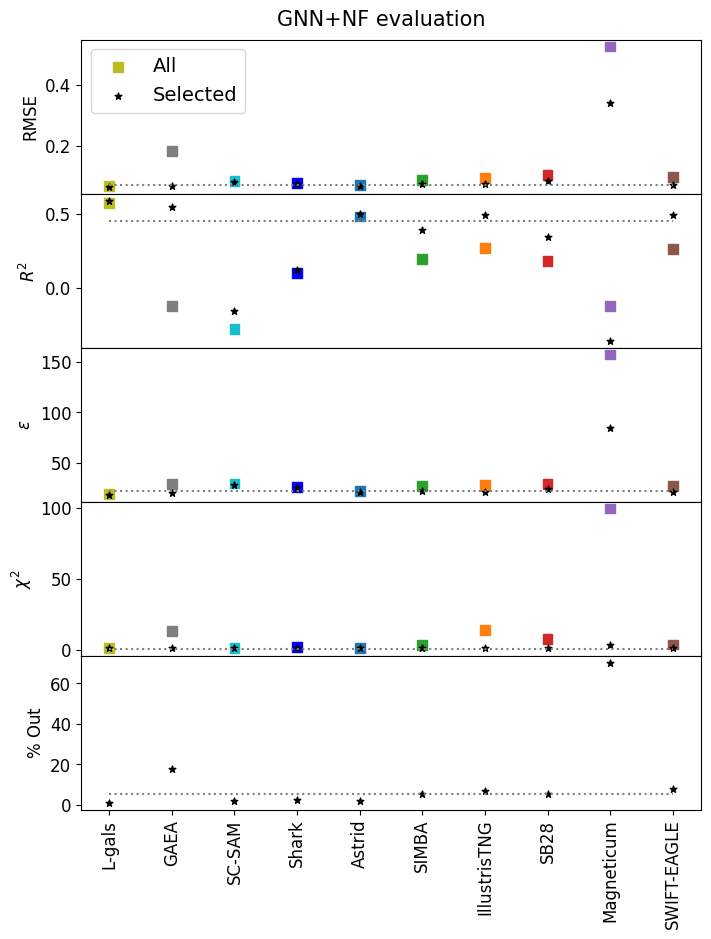}
    \caption{{\bf GNN+NF: Model Evaluation.} We present RMSE (root mean square error), $R^2$ (coefficient of determination), $\epsilon$ (mean relative error), $\chi^2$ (reduced chi squared), and $\%$ Out (percentage of samples removed) for the test sets of L-Galaxies, GAEA, SC-SAM, IllustrisTNG, {\sc SIMBA}, SB28, Magneticum, and SWIFT-EAGLE indicated in the x-axis. Results are presented to all and selected samples according to $\chi^2$ values. \label{fig:metrics-GNN+NF}}
% \end{figure}
\end{wrapfigure}

In this work we have been using $p (\mathbf{x})$ as a Gaussian distribution and conditioning the latent space from the GNNs to an Autoregressive Spline \citep{NeuroSplines2019, Dolatabadi2020, auto_regressive-2016} to infer $\Omega_m$ distributions.
The implementation was performed using \href{https://docs.pyro.ai/en/stable/index.html}{Pyro} library \citep{bingham2019pyro}.

The graph construction, GNN architecture, and training procedure followed the same scheme described in the main part of the paper (see Section \ref{sec:methodology}).
The primary difference lies in the number of catalogs used for training, validation, and testing, which are $600$, $200$, and $200$, respectively.
In addition, the optimal set of hyperparameters was determined using the {\em Optuna} optimization package \citep{optuna_2019}, with the selected values listed in Table \ref{tab:hyperparametersGNN+NF}.
The hyperparameters associated with the NF are bound and count bins, which respectively represent the number of spline segments and the size of the bounding box used in the transformation.

To facilitate a direct comparison between this model and the main one presented in the paper, we computed the mean and standard deviation of each posterior prediction for $\Omega_{\rm m}$.
The results for the GNN coupled with the NF are shown in Figures \ref{fig:result-GNN+NF} and \ref{fig:metrics-GNN+NF}, alongside the corresponding results for the baseline model in Figures \ref{fig:main_result} and \ref{fig:metrics_main}.

The takeaway from Figure \ref{fig:result-GNN+NF} is that, overall, the predictions exhibit a slight bias and larger uncertainties, while compared to the main model.
These biases are particularly noticeable for the CV sets of {\sc Shark}, \texttt{Astrid}, and Magneticum.
It is also important to note that the $\chi^2$ values increased across several models, reaching $13.5$ for GAEA, approximately $2$ for {\sc Shark}, and even higher values for the hydrodynamical simulations ($13.8$ for IllustrisTNG and around $99$ for Magneticum).

\begin{table*}[h!]
 \caption{\label{tab:hyperparametersGNN+NF} Hyperparameters values selected for the best model.}
 \begin{center}
% \resizebox{\textwidth}{!}{%
  \begin{tabular}{cc}
   \hline\hline
   \textbf{Hyperparameter} & \textbf{Best value} \\
   \hline\hline
   Bound & 3\\
   Count Bins & 9\\
   Hidden Features & 94\\
   Number of GNN and NF layers & 5\\
   Number of NF transformations & 4\\
    $r_{link}$ & $0.25$Mpc/h\\
    Learning Rate & $8.76 \cdot 10^{-4}$\\
    Weight Decay & $7.4 \cdot 10^{-4}$\\
   \cline{1-2}
   \end{tabular}%}
  \end{center}
\end{table*}

This increase in $\chi^2$ propagated into the fraction of predictions excluded under the $\chi^2 > 10$ criterion.
Among the SAMs, $17.6 \%$ of the GAEA catalogs were removed, while for the Magneticum hydrodynamical simulation this fraction increased to $70.1 \%$.
Nevertheless, after excluding these outlier catalogs, the remaining $\chi^2$ values stabilized around unity for all tests, indicating good performance on the retained samples.

The GNN+NF model evaluation summarized in Figure \ref{fig:metrics-GNN+NF} also reveals that it performs slightly worse than the main GNN+MNN model.
On average, the RMSE increased to about $0.07$, the $R^2$ values stabilized near $0.45$, the $\chi^2$ remained close to $1$, and the fraction of outliers increased to roughly $10 \%$.

The final evaluation of the generative model is performed through a coverage test, a metric used to assess the accuracy of generative posterior estimators, $\hat{p}(\Omega_{\rm m} | G)$.
Here, we employ the Tests of Accuracy with Random Points (TARP) \citep{lemos2023}.

The basic idea of this test is to compute the expected coverage probability (ECP), denoted as $\mathrm{ECP}(\hat{p}, \alpha, \mathcal{R})$, averaged over the data distribution $G$, where $\hat{p}$ is the posterior estimator, $\mathcal{R}$ is a credible region generator, and $\alpha$ represents the credibility level.
The ECP is estimated using samples drawn from both the reference distribution $p(\Omega_{\rm m} | G)$, and the inferred posterior $\hat{p}(\Omega_{\rm m} | G)$.

A high-quality posterior estimator should yield a one-to-one correspondence between coverage and credibility level across the full range of $\Omega_{\rm m}$.
Deviations from this relation indicate that the posterior estimator is either overconfident (undercovering) or underconfident (overcovering).

\begin{figure}[h!]
 \centering
 \includegraphics[scale=0.47]{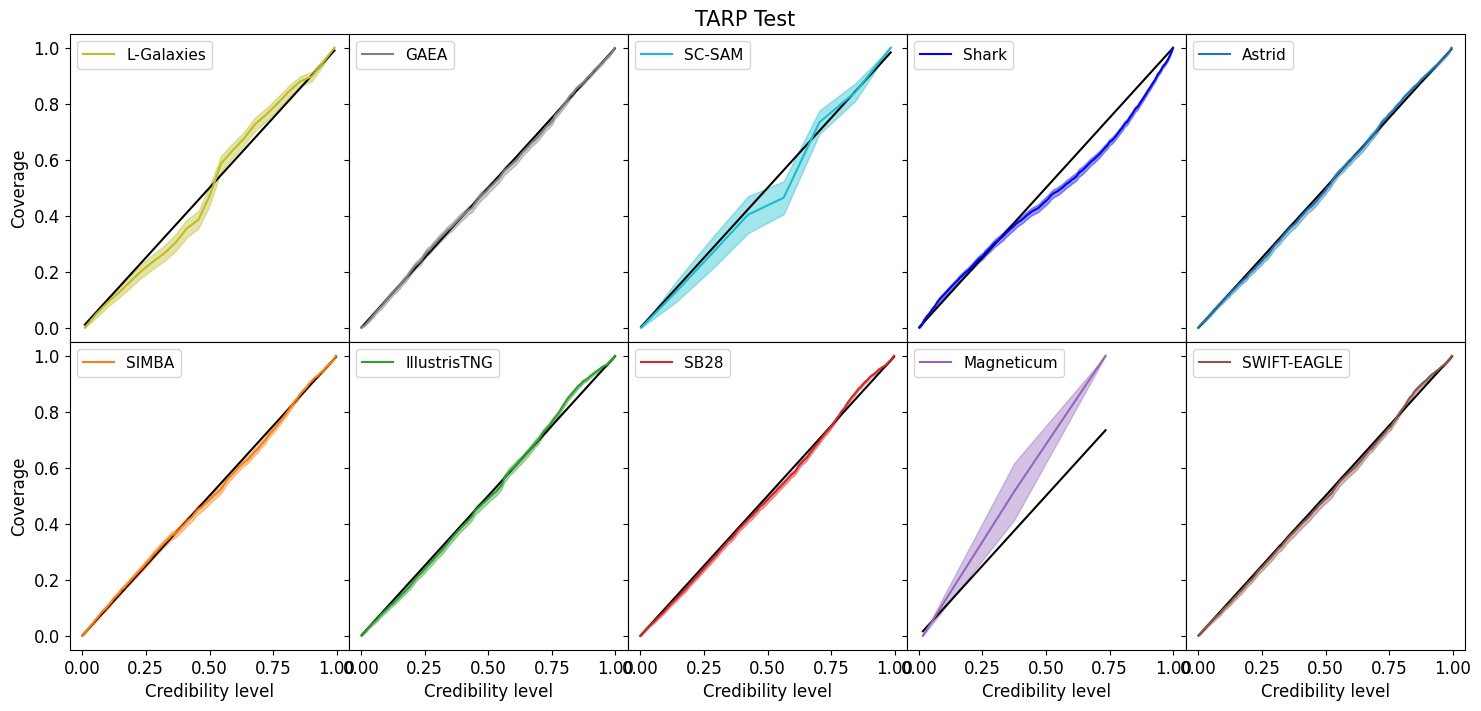}
 \caption{{\bf GNN+NF: TARP test.} The TARP test was performed for the testing sets of L-Galaxies, GAEA, SC-SAM, {\sc Shark}, \texttt{Astrid}, {\sc SIMBA}, IllustrisTNG, SB28, Magneticum, and SWIFT-EAGLE catalogs. Here we present the coverage versus the credibility level for each dataset.}
 \label{fig:TARP}
\end{figure}

The results of the TARP test are shown in Figure \ref{fig:TARP}, where the solid lines represent the mean values and the shaded regions correspond to the standard deviation across $100$ realizations of the test.
We present the results for the testing sets of L-Galaxies, GAEA, SC-SAM, {\sc Shark}, \texttt{Astrid}, {\sc SIMBA}, IllustrisTNG, SB28, Magneticum, and SWIFT-EAGLE.

The model's quality can be most directly evaluated from the L-Galaxies test, which already shows a slight deviation from the one-to-one correspondence line, indicating a transition from mild overconfidence to underconfidence.
However, this result should be interpreted with caution, as the test set includes only $200$ samples of $\Omega_{\rm m}$, which may be insufficient for a reliable coverage assessment.
Because the total number of available L-Galaxies realizations is $1,000$, the maximum number of independent test samples that could be used was necessarily limited to $200$.
A similar limitation may affect the results for SC-SAM and Magneticum, which include only $73$ and $77$ catalogs, respectively.

Apart from the {\sc Shark} test, where the model appears slightly overconfident at credibility levels above $30 \%$, the TARP results for GAEA, \texttt{Astrid}, {\sc SIMBA}, IllustrisTNG, SB28, and SWIFT-EAGLE show an almost perfect correspondence between coverage and credibility, demonstrating the robustness of the generative model across these datasets.

%% For this sample we use BibTeX plus aasjournals.bst to generate the
%% the bibliography. The sample631.bib file was populated from ADS. To
%% get the citations to show in the compiled file do the following:
%%
%% pdflatex sample631.tex
%% bibtext sample631
%% pdflatex sample631.tex
%% pdflatex sample631.tex

\bibliography{sample631}{}
\bibliographystyle{aasjournal}

%% This command is needed to show the entire author+affiliation list when
%% the collaboration and author truncation commands are used.  It has to
%% go at the end of the manuscript.
%\allauthors

%% Include this line if you are using the \added, \replaced, \deleted
%% commands to see a summary list of all changes at the end of the article.
%\listofchanges

\end{document}